\documentclass[aps,prd,preprint,preprintnumbers,showpacs,showkeys,nofootinbib,superscriptaddress]{revtex4-1}

\usepackage{booktabs}    
\usepackage{placeins}
\usepackage{soul}
\usepackage[usenames,dvipsnames,svgnames]{xcolor}
\definecolor{redd}{rgb}{0.8, 0.1,0.2}
\definecolor{navy}{rgb}{0.05, 0.23,0.75}
\usepackage[colorlinks=true,citecolor=red,linkcolor=blue]{hyperref}
\usepackage{chngcntr}
\hypersetup{
     colorlinks   = true,
     citecolor    = navy,
	linkcolor = redd,
	urlcolor=navy,
	anchorcolor=blue
}
\usepackage{bbm}
\usepackage{amsfonts}
\usepackage{amsmath,amssymb}
\usepackage{mathrsfs}
\usepackage{epsfig}
\usepackage{graphicx}
\usepackage{wrapfig}                 
\usepackage{url}
\usepackage{float}
\usepackage{pstricks}
\usepackage{color}
\usepackage{multirow}
\usepackage{lipsum}
\usepackage{enumitem}
\usepackage{ctable}
\newcolumntype{L}{>{\centering\arraybackslash}m{1.5cm}}

\usepackage[normalem]{ulem}

\usepackage{enumitem}

\usepackage{chngcntr}

\newcommand{\bear}{\begin{array}}
\newcommand {\eear}{\end{array}}

\newcommand{\beq}{\begin {equation}}
\newcommand{\eeq}{\end   {equation}}
\newcommand{\bea}{\begin {eqnarray}}
\newcommand{\eea}{\end   {eqnarray}}
\newcommand{\baa}{\begin {array}   }
\newcommand{\eaa}{\end   {array}   }
\newcommand{\bit}{\begin {itemize} }
\newcommand{\eit}{\end   {itemize} }
\newcommand{\be }{\begin {equation}}
\newcommand{\ee }{\end   {equation}}
\newcommand{\nn }{\nonumber        }

\def\bea{\begin{eqnarray}}
\def\eea{\end{eqnarray}}

\newcommand{\bef}{\begin{figure}}
\newcommand {\eef}{\end{figure}}
\newcommand{\bec}{\begin{center}}
\newcommand {\eec}{\end{center}}

\definecolor{cerulean}{rgb}{0., 0.62,0.7}

\newcommand{\twiddle}{{\raise.17ex\hbox{$\scriptstyle\sim$}}}

\begin{document}

\title{\Large Holographic Completion of Minimal Neutral Naturalness Model and Deconstruction} 

\author{Ling-Xiao Xu}
\email{lingxiaoxu@pku.edu.cn}
\affiliation{Department of Physics and State Key Laboratory of Nuclear Physics and Technology, 
Peking University, Beijing 100871, China}

\author{Jiang-Hao Yu}
\email{jhyu@itp.ac.cn}
\affiliation{CAS Key Laboratory of Theoretical Physics, Institute of Theoretical Physics, Chinese Academy of Sciences, Beijing 100190, P. R. China}
\affiliation{School of Physical Sciences, University of Chinese Academy of Sciences, No.19A Yuquan Road, Beijing 100049, P.R. China}

\author{Shou-hua Zhu}
\email{shzhu@pku.edu.cn}
\affiliation{Department of Physics and State Key Laboratory of Nuclear Physics and Technology, 
Peking University, Beijing 100871, China}
\affiliation{Collaborative Innovation Center of Quantum Matter, Beijing, 100871, China}
\affiliation{Center for High Energy Physics, Peking University, Beijing 100871, China}

\begin{abstract}
We previously presented the ``minimal neutral naturalness model", in which neutral naturalness is realized with the minimal coset and the minimal field content in the color-neutral sector. To fully eliminating the cutoff dependence in the Higgs potential, its ultraviolet completion was laid out. In this work, we investigate in details the holographic realization in warped five-dimensional framework. Using the holographic method, we obtain the effective action on the UV brane and derive the finite Higgs potential, with the Higgs being a pseudo Nambu-Goldstone boson. The Higgs potential is fully radiatively generated and vacuum misalignment is naturally realized through only the fermionic contribution. To illustrate general features of the lowest Kaluza-Klein states, we construct the corresponding deconstructed two-site composite model.

\end{abstract}

\maketitle

\section{Introduction}
\label{sec:intro}
The discovery of the Higgs boson marked a milestone in the history of particle physics. As the agent of electroweak symmetry breaking (EWSB), the Higgs boson plays the central role of giving masses to all the particles in the standard model (SM). In the SM, the Higgs boson $h$ is assumed as an elementary scalar arising from a $SU(2)_L$ doublet $H=\frac{1}{\sqrt{2}}(w_1+i w_2, h+i w_3)^T$, with the Goldstone bosons $w_{1,2,3}$ being eaten by electroweak gauge bosons after EWSB, and the origin of EWSB is described by the so-called Ginzburg-Landau potential. However, since we have not measured shape of the Higgs potential yet, we do not know whether the Higgs boson is an elementary particle and the potential is Ginzburg-Landau-like. Most generally, from a bottom-up perspective, one can naively add higher dimensional operators (e.g.~\cite{Buchmuller:1985jz,Grzadkowski:2010es,Giudice:2007fh}) to describe possible deviations from the SM,
\bea
V(H)=-\mu^2 H^\dagger H+\lambda (H^\dagger H)^2+\frac{c_6}{\Lambda^2} (H^\dagger H)^3+\cdots
\label{eq1}
\eea
where $\mu^2,\lambda$ are the ordinary coefficients for the Higgs quadratic and quartic terms respectively, and $\Lambda$ is some new physics scale beyond the electroweak scale $v=246$ GeV, $c_6$ is the corresponding Wilson coefficient of the dimension-six operator $(H^\dagger H)^3$.

\begin{figure}
\includegraphics[scale=0.35]{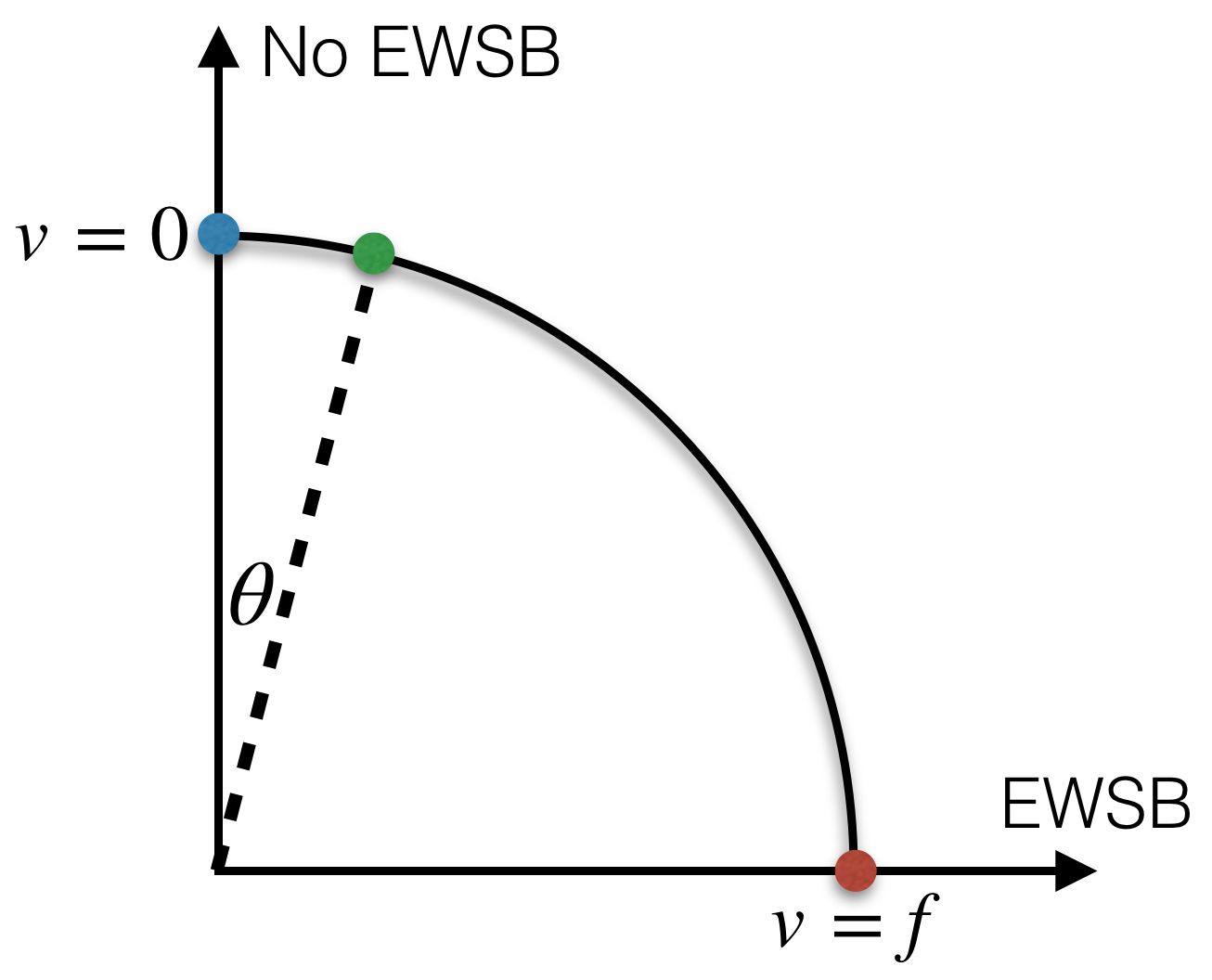}
\caption{Sketch of vacuum misalignment for the case that the Higgs boson is assumed as a PNGB. The circle denote the vacuum of global symmetry breaking with the symmetry breaking scale $f$. The dots in blue, green, and red denote the vacuum at which the electroweak scale $v=0$, $v<<f$, and $v=f$, respectively. Note that the precision Higgs data prefers the vacuum in green ($v<<f$).}
\label{vac_mis}
\end{figure}

Instead of assuming the fundamental nature of the Higgs boson as in SM, one of the most attractive and theoretical-motivated alternatives is to assume it as a pseudo Nambu-Goldstone boson (PNGB); see Refs.~\cite{Bellazzini:2014yua, Panico:2015jxa} for general reviews. In this case, the Higgs potential is schematically
\bea
V(H)\simeq -\gamma \sin^2\left(\frac{\sqrt{H^\dagger H}}{f}\right)+\beta  \sin^4\left(\frac{\sqrt{H^\dagger H}}{f}\right)
\label{pot1}
\eea
considering its nature as a PNGB and a $SU(2)_L$ doublet, where $f$ is the global symmetry breaking scale at which the PNGB Higgs emerges, while $\gamma$ and $\beta$ are two coefficients dictated by the dynamics responsible for generating the Higgs potential. When $v\ll f$, one can expand the Higgs potential in Eq.~\ref{pot1} in powers of $H^\dagger H$ and match to the effective Lagrangian in Eq.~\ref{eq1}. By assuming the Goldstone nature, the so-called hierarchy problem can be solved with some symmetries (as well as top partners) being introduced to cut off the quadratic divergence, i.e., the Higgs mass is not quadratically sensitive to higher scale~\cite{ArkaniHamed:2001ca}. Furthermore, due to the ``shift symmetry", the above Higgs potential in Eq.~\ref{pot1} is usually generated at loop level and its specific form is controlled by some explicit shift-symmetry-breaking effects. Given the Higgs potential in Eq.~\ref{pot1}, there are two steps of symmetry breaking, i.e., the global symmetry breaking and the electroweak symmetry breaking. The vacuum structure is depicted schematically in Fig.~\ref{vac_mis}.
One can define a so-called ``vacuum misalignment"~\cite{Kaplan:1983fs} angle to denote the magnitude of misalignment of these two scales, i.e.,
\bea
\xi\equiv\sin^2\theta= \frac{v^2}{f^2}=\frac{\gamma}{2\beta}\ .
\eea
Note that, by minimizing the Higgs potential, the misalignment angle is related to the dynamics that generates the potential.
Note that the current data of Higgs precision measurements already set a bound on the ratio of these two symmetry breaking scales, i.e., $\theta\ll 1$~\cite{Li:2019ghf}. This leads to a bound on the ratio of the coefficients $\gamma$ and $\beta$,
\bea
\frac{\gamma}{2\beta}\ll 1.
\eea
Unfortunately, it is not straightforward to realize the above vacuum misalignment condition theoretically. It typically relies on cancellation between fermionic and bosonic contributions to the $\gamma$ and $\beta$ in the radiative Higgs potential, to realize a small vacuum misalignment angle, because the fermionic contribution to the potential usually gives $v\simeq f$, while the bosonic contribution tends to preserve the electroweak symmetry $v=0$. Due to the compensation between these two effects of fermions and bosons, the vacuum with $v\ll f$ is realized. 

Instead of considering the compensation between fermionic and bosonic contribution, it is highly motivated to develop a strategy to realize the vacuum  
$v\ll f$ within only the fermionic sector. 
Such a model construction would naturally realize the misalignment, because there is no need of realizing the compensation between two different sectors. In Ref.~\cite{Xu:2018ofw}, it has been shown that this scenario can be realized in composite Higgs model with the minimal coset $SO(5)/SO(4)$~\cite{Agashe:2004rs,Contino:2006qr}, from which the Higgs boson arises as a PNGB. Explicitly, vacuum misalignment is realized in the fermionic sector considering both the top quark and the top partners in the color-neutral sector. Because the top partners do not have QCD quantum number and thus cannot be produced with large rate at the Large Hadron Collider (LHC), light top partners with sub-TeV mass are still viable options and the little hierarchy problem is avoided. This kind of construction can be regarded as one of the incarnations of the neutral naturalness scenario~\cite{Chacko:2005pe, Cai:2008au, Craig:2014aea, Geller:2014kta, Barbieri:2015lqa, Low:2015nqa, Burdman:2006tz,Cohen:2015gaa,Yu:2016bku,Yu:2016swa,Yu:2016cdr,Badziak:2017syq,Badziak:2017kjk,Badziak:2017wxn, Serra:2017poj, Csaki:2017jby, Dillon:2018wye, Cohen:2018mgv,Cheng:2018gvu}. Compared to other model realizations, the model in Ref.~\cite{Xu:2018ofw} is characterized by the minimal coset from which the PNGB Higgs emerges, and the minimal field content in the color-neutral sector for modeling the realistic EWSB, therefore it is named as the minimal neutral naturalness model (MNNM).

The $\gamma$ and $\beta$ in the radiative Higgs potential typically have logarithmic dependence on the cutoff scale, which comes from some unknown ultraviolet (UV) physics.
To remove the log divergence and obtain the finite Higgs potential, we construct the holographic completion of MNNM and the corresponding deconstructed version. In the five-dimensional holographic Higgs setup, the PNGB Higgs is treated as the fifth component of five-dimensional gauge field $A_5(x_\mu,y)$ living in the bulk~\cite{Contino:2003ve}. 
Because of the compositeness being introduced at higher scale, the holographic/deconstructed setup can be viewed as one of the possible UV completions of the minimal model. With the standard holographic technique (see Refs.~\cite{Serone:2009kf,Contino:2010rs} for pedagogical reviews and references therein), the contribution from all the composite resonances to the Higgs potential can be resummed, whose information is then encoded in the form factors after these resonances being integrated out. Furthermore, the contribution from the composite resonances renders the finite Higgs potential, which means the Higgs mass (and the electroweak scale) is insensitive to the UV cutoff. With the assumption that there is no mass splitting inside the full composite multiplets, the Higgs potential can also be finite in the deconstructed two-site model.

The rest of this paper is organized as follows. In Sec.~\ref{sec:model}, the basic five-dimensional model setup is presented; the holographic technique is applied to the model in Sec.~\ref{sec:holography}; the effective holographic Lagrangian is obtained, and the Higgs potential is calculated in Sec.~\ref{sec:potential}; Constraints from the electroweak precision tests are briefly discussed in Sec.~\ref{sec:precision}; in Sec.~\ref{sec:deconstruction}, we present in details the deconstructed version of the five-dimensional model, then followed by comments on the finiteness of the Higgs potential in the deconstructed model; finally we conclude in Sec.~\ref{sec:conclusion}.

\section{The Five-Dimensional Model}
\label{sec:model}
The five dimensional theory can be defined in the interval $x_5\in [0,\pi R]$ with the geometry given by the metric
\bea
ds^2\equiv a^2(x_5)\eta_{\mu\nu} dx^\mu dx^\nu-dx_5^2\ , 
\label{metric1}
\eea
where the UV brane and the IR brane are located at $x_5=0$ and $x_5=\pi R$ respectively, and $\eta_{\mu\nu}=\text{diag}(1, -1, -1, -1)$ denote the usual four-dimensional metric. For five-dimensional anti-de Sitter space $AdS_5$~\cite{Randall:1999ee}, we explicitly have the curvature factor
$a^2(x_5)=e^{-2kx_5}$.
Equivalently, one can also use the following metric to denote the $AdS_5$,
\bea
ds^2\equiv a(z)^2\ (\eta_{\mu\nu}dx^\mu dx^\nu-dz^2)=\left(\frac{L}{z}\right)^2(\eta_{\mu\nu}dx^\mu dx^\nu-dz^2)
\label{metric2}
\eea
where the theory is instead defined $z_{UV}\leq z\leq z_{IR}$ with the UV brane localized at $z_{UV}=L_0\sim 1/M_{\text{planck}}$, the IR brane localized at $z_{IR}=L_1\sim 1/\text{TeV}$, and the $AdS$ curvature radius denoted by $L$ accordingly. The above two notations can be translated to each other by the redefinition of coordinates, i.e., $L/z=e^{-kx_5}$ where $L=1/k$. 

\begin{figure}
\includegraphics[scale=0.35]{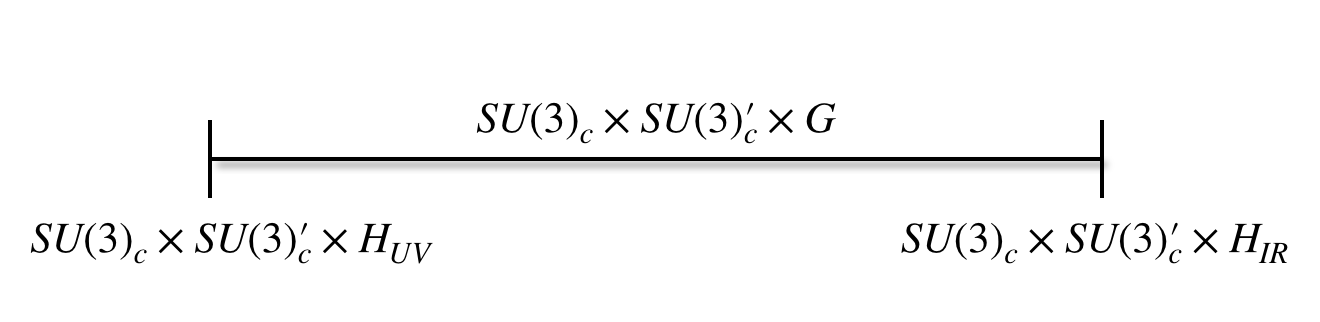}\\
\includegraphics[scale=0.3]{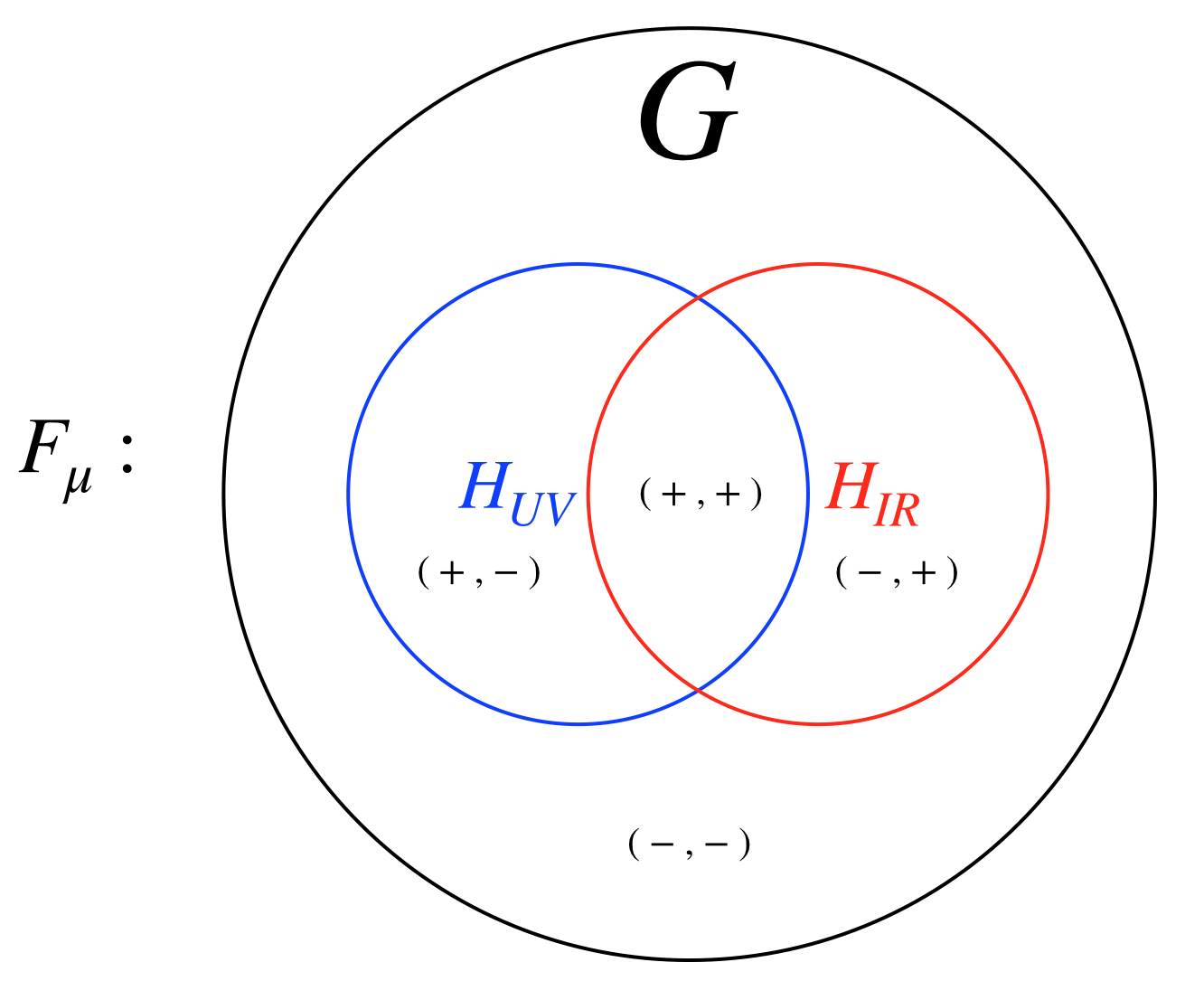}
\caption{The symmetry breaking pattern and the corresponding boundary conditions for the five-dimensional gauge theory. The black circle denotes the bulk symmetry, while the blue circle and the red circle denotes the symmetry on the UV brane and IR brane, respectively.}
\label{sym_brk}
\end{figure}

After introducing the metric, the action for the five-dimensional gauge theory can be written as 
\bea
S_{gauge}=\int \text{d}^4x\int\text{d}x_5\ \sqrt{g}\ \left\{\sum_b -\frac{1}{4g_b^2} \mathcal{F}^{bMN} \mathcal{F}^b_{MN}\right\}
\eea
where the index $b$ denotes the all internal degrees of freedom of the gauge bosons living in the bulk, and $M,N$ denote the five-dimensional spacetime coordinates ($M, N=0,1,2,3,4$). Considering the internal components of the gauge field $F^b_M$, the bulk gauge symmetry $G$ needs to be broken down to $H_{UV}$ and $H_{IR}$ at the UV and IR brane, respectively. In practice, the symmetry breaking can be realized with the following Neumann ($+$) and Dirichlet ($-$) boundary conditions (B.C.) imposed on the above two branes,
\begin{align}
\partial_5 F^{A}_\mu (x_5=0,\pi R)= F^{A}_5 (x_5=0,\pi R)=0\ \ \ \ \ &A\in H_{UV}\cap H_{IR};\nn\\
\partial_5 F^{\dot{A}}_\mu (x_5=0)= F^{\dot{A}}_5 (x_5=0)=0,\ F^{\dot{A}}_\mu (x_5=\pi R)= \partial_5 F^{\dot{A}}_5 (x_5=\pi R)=0\ \ \ &\dot{A}\in H_{UV}/(H_{UV}\cap H_{IR});\nn\\
\partial_5 F^{\tilde{A}}_5 (x_5=0)= F^{\tilde{A}}_\mu (x_5=0)=0,\ F^{\tilde{A}}_5 (x_5=\pi R)= \partial_5 F^{\tilde{A}}_\mu (x_5=\pi R)=0\ \ \ &\tilde{A}\in H_{IR}/(H_{UV}\cap H_{IR});\nn\\
\partial_5 F^{\bar{A}}_5 (x_5=0,\pi R)= F^{\bar{A}}_\mu (x_5=0,\pi R)=0\ \ \ \ \ &\bar{A}\in G/(H_{UV}\cup H_{IR}),
\end{align}
where the internal degrees of freedom are decomposed as $b= \{A, \dot{A}, \tilde{A}, \bar{A}\}$.
More straightforwardly, we sketch the above symmetry breaking pattern and the associated B.C. for the four-dimensional gauge fields $F^b_\mu$ in Fig.~\ref{sym_brk}. Note that the corresponding fifth component $F^b_5$ always has the opposite B.C. with respect to $F^b_\mu$.

For realistic composite Higgs models, symmetries are $H_{UV}=SU(2)_L\times U(1)_Y$, $H_{IR}=SO(4)\times U(1)_X$ and $G=SO(5)\times U(1)_X$~\cite{Agashe:2004rs}, in addition to the gauge group $SU(3)_c$ for the QCD color and possibly another $SU(3)^\prime_c$ for the color in the hidden sector. As we will see explicitly in the fermionic sector, only the top partners in the hidden sector are charged under this $SU(3)^\prime_c$. Since both the groups $SU(3)_c$ and $SU(3)^\prime_c$ are irrelevant to symmetry breaking, we will neglect them for a moment. On the other hand, the unbroken $SO(4)$ symmetry on the IR brane can serve as the custodial symmetry under which the $\rho$ parameter is protected. Accordingly, we see that the SM gauge symmetry $SU(2)_L\times U(1)_Y$ is unbroken on the UV brane, and it is embedded in $SO(4)\times U(1)_X\cong SU(2)_L\times SU(2)_R\times U(1)_X$, with the hyper-charge being identified as $Y=X+T^3_R$. As we can see, this additional $U(1)_X$ here is necessary for obtaining the correct hyper-charges for the SM fermions.

Following the standard approach of Kaluza-Klein (KK) decomposition, we can expand the gauge fields as
\bea
F_\mu^a(x,x_5)=\sum_n f_n^a(x_5) F_\mu^{n,a}(x), \ \ \ F_5^a(x,x_5)=\sum_n \tilde{f}_n^a(x_5) F_5^{n,a}(x)
\eea
where the solutions of KK profiles $f_n^a(x_5)$ and $\tilde{f}_n^a(x_5)$ follow the equation
\bea
\partial_5(a^2(x_5)\partial_5f_n)+m_n^2f_n=0
\eea
and the normalization condition $g_5^2\int_0^Lf_nf_m=\delta_{nm}$. And $F_\mu^{n,a}(x)$ and $F_5^{n,a}(x)$ are the four-dimensional fields after KK decomposition accordingly. Let us denote the two independent solutions as $C(x_5,m_n)$ and $S(x_5,m_n)$, with their explicit forms given by
\begin{align}
C(x_5,m_n)&=\frac{\pi m_n}{2k}a^{-1}(x_5)\left[Y_0\left(\frac{m_n}{k}\right)J_1\left(\frac{m_n}{ka}\right)-J_0\left(\frac{m_n}{k}\right)Y_1\left(\frac{m_n}{ka}\right)\right],\\
S(x_5,m_n)&=\frac{\pi m_n}{2k}a^{-1}(x_5)\left[J_1\left(\frac{m_n}{k}\right)Y_1\left(\frac{m_n}{ka}\right)-Y_1\left(\frac{m_n}{k}\right)J_1\left(\frac{m_n}{ka}\right)\right].
\end{align}
After KK decomposition, we see only the fields with the ($+,+$) boundary conditions, i.e., the field components $F^{A}_\mu$ and $F^{\bar{A}}_5$, have the zero modes
\bea
S_{gauge}=\int \text{d}^4x \left\{\frac{1}{2}\left(\partial_\mu F^{\bar{A}}_5\right)^2-\frac{1}{4} F^A_{\mu\nu}F^{A\mu\nu}+\text{massive KK modes}\right\}.
\eea
At the tree level, we see that $F^{\bar{A}}_5$ is nothing but a massless scalar from a four-dimensional perspective. Due to quantum correction, it can be identified as the PNGB Higgs of the coset $SO(5)/SO(4)$ and can be dubbed as $h(x)$ for convenience. The corresponding Wilson line of $F^{\bar{A}}_5$ is then
\bea
\Omega(x)=e^{-i h(x) \hat{T} g_5 \left(\int_0^{L}\text{d}y\ a^{-2}\right)^{1/2}}
\label{wilsonline}
\eea
with the ``decay constant" of the PNGB Higgs identified as $f^{-2}=g_5^2\int_0^L\text{d}y\ a^{-2}$, and $g_5$ the $5$D gauge coupling with dimension $\text{Dim}[1/g_5^2]=1$. Note that one can make use of Eq.~\ref{wilsonline} to remove the $A_5$ dependence in the fifth dimension via certain gauge transformations; see e.g. Ref.~\cite{Panico:2007qd}.
One can also obtain the mass spectrum of gauge bosons after EWSB, i.e., when the Higgs boson develops a non-zero vacuum expectation value (VEV), $h=\langle h\rangle$~\cite{Falkowski:2006vi,Medina:2007hz}.  

In Ref.~\cite{Xu:2018ofw}, two pairs of vector-like top partners, namely one $SU(2)_L$ doublet and one $SU(2)_L$ singlet, are introduced to cut off the quadratic divergence of the Higgs mass and at the same time realize the natural vacuum misalignment. In the five-dimensional setup, these fermions can be embedded in full representations of the bulk gauge symmetry $SO(5)$. For our purpose, we introduce two $SO(5)$ fundamental $5$-plets $\xi_q$ and $\xi_{\widetilde{q}}$, and two $SO(5)$ singlets $\xi_t$ and $\xi_{\widetilde{T}}$ living in the bulk. The fields $\xi_q$ and $\xi_t$ are only charged under the QCD $SU(3)_c$ color group, while the fields $\xi_{\widetilde{q}}$ and $\xi_{\widetilde{T}}$ are QCD neutral but charged under the $SU(3)^\prime_c$ color in the hidden sector.
We have the Lagrangian for the bulk fermions as  
\begin{align}
S_{fermion}&=\int \text{d}^4x\int_{L_0}^{L_1}\text{d}z \sqrt{g}\left\{\sum_{f={q,t,\widetilde{q},\widetilde{T}}}\frac{1}{g_5^2}\left(\frac{i}{2}\bar{\xi_f}\gamma^M D_M\xi_f-\frac{i}{2}\left(D_M\xi_f\right)^\dagger\Gamma^0\gamma^M\xi_f-M_{\xi_f}\bar{\xi_f}\xi_f\right)\right\}\ ,
\end{align}
with the specific B.C. assignments on each component of the bulk fermions as
\begin{align} 
\xi_q&=\left[
\baa{cc}
(2,2)^q_L=\left[\baa{c}q^\prime_L(-+)\\q_L(++)\eaa\right]& (2,2)^q_R=\left[\baa{c}q^\prime_R(+-)\\q_R(--)\eaa\right]\\
(1,1)^q_L(-+)&(1,1)^q_R(+-)
\eaa
\right],\nn\\
\xi_t&=\left[\baa{cc}(1,1)^t_L(--)&(1,1)^t_R(++)\eaa\right];
\label{bulksm}
\end{align}
and 
\begin{align}
\xi_{\widetilde{q}}&=\left[
\baa{cc}
(2,2)^{\widetilde{q}}_L=\left[\baa{c}\widetilde{q}^\prime_L(-+)\\ \widetilde{q}_L(++)\eaa\right]& (2,2)^{\widetilde{q}}_R=\left[\baa{c}\widetilde{q}^\prime_R(+-)\\ \widetilde{q}_R(--)\eaa\right] \\
(1,1)^{\widetilde{q}}_L(++)&(1,1)^{\widetilde{q}}_R(--)
\eaa
\right],\nn\\
\xi_{\widetilde{T}}&=\left[\baa{cc}(1,1)^{\widetilde{T}}_L(--)&(1,1)^{\widetilde{T}}_R(++)\eaa\right],
\label{bulkneutral}
\end{align}
where the chiralities under the four-dimensional Lorentz group are denoted by $L$ and $R$, respectively. The notations $(1,1)$ and $(2, 2)$ denote the quantum numbers under the $SU(2)_L\times SU(2)_R$ symmetry when decomposing a full $SO(5)$ representation into $SO(4)$ representations, as $5=4+1=(2,2)+(1,1)$ in our model.

As we will see later, brane-localized terms are needed to explicitly break the ``shift symmetry" of the Higgs boson; otherwise, rather than being a pseudo Nambu-Goldstone boson, the Higgs boson would have become an exact Nambu-Goldstone boson, without non-derivative terms such as Higgs potential and top Yukawa coupling. 
Specifically, for the fermions charged under the QCD $SU(3)_c$ color symmetry, we introduce the term
\begin{align}
S_{mixing}=\ &\frac{m}{g_5^2}\int\text{d}^4x\sqrt{-g_{\text{ind}}}\ \overline{(1,1)^q_L}\ (1,1)^t_R\ (z_{IR}=L_1)+\text{h.c.}\nn\\
\equiv\ &\frac{m}{g_5^2}\int\text{d}^4x\sqrt{-g_{\text{ind}}}\ \overline{\xi^{(1)}_{qL}}\ \xi_{tR}\ (z_{IR}=L_1)+\text{h.c.}\ ,
\label{IR1}
\end{align}
on the IR brane, which is invariant under $SO(4)\times U(1)_X$. Note that $\xi_q$ is decomposed as $\xi_q\equiv (\xi^{(4)}_q,\xi^{(1)}_q)^T$ under the $SO(4)$ symmetry, $m$ is the dimensionless mixing parameter. With our conventions, all the bulk fermions are of the dimension $\text{Dim}[\xi]=3/2$.
For the fermions only charged under the $SU(3)^\prime_c$ group, there is similarly a mixing term
\begin{align}
S^\prime_{mixing}=\ &\frac{\widetilde{m}}{g_5^2}\int\text{d}^4x\sqrt{-g_{\text{ind}}}\ \overline{(1,1)^{\widetilde{q}}_L}\ (1,1)^{\widetilde{T}}_R\ (z_{IR}=L_1)+\text{h.c.}\nn\\
\equiv\ &\frac{\widetilde{m}}{g_5^2}\int\text{d}^4x\sqrt{-g_{\text{ind}}}\ \overline{\xi^{(1)}_{\widetilde{q}L}}\ \xi_{\widetilde{T}R}\ (z_{IR}=L_1)+\text{h.c.}\ ,
\label{IR2}
\end{align}
with the fermions in the hidden sector also being decomposed according to the $SO(4)$ symmetry.
As we mentioned earlier, the form of the radiative Higgs potential is determined by the explicit global-symmetry-breaking effects; in the current case, the Higgs potential depends on mixing parameters $m$ and $\widetilde{m}$ in Eq.~\ref{IR1} and Eq.~\ref{IR2}. Furthermore, we need to introduce the right-handed component $\widetilde{q}_{0R}$, for making the doublet top partner in the hidden sector become vector-like and uplifting the zero mode accordingly. For that, we can explicitly have the mass term for the doublet fermion on the UV brane
\bea
S_{mass}=-\frac{\widetilde{m}_q}{g_5^2}\int\text{d}^4x\sqrt{-g_{\text{ind}}}\ \overline{\widetilde{q}}_{0R}\ \widetilde{q}_L(++)\ (z_{UV}=L_0)+\text{h.c.}\ ,
\label{mass}
\eea
which is invariant under $SU(2)_L\times U(1)_Y$. Note that the minus sign is just conventional and it is defined as above for later convenience.

One can define the KK profiles for the bulk fermions in general as
\begin{align}
\Psi_L(x,x_5)&=a^{-2}e^{M_fx_5}\sum_n f_{L,n}(x_5)\Psi_{L,n}(x),\\
\Psi_R(x,x_5)&=a^{-2}e^{-M_fx_5}\sum_n f_{R,n}(x_5)\Psi_{R,n}(x),
\end{align}
where $\Psi_{L,n}(x)$ and $\Psi_{R,n}(x)$ are the fermionic field after KK decomposition of the fermionic fields with $n$ denotes the corresponding KK state.
The KK profiles $f_{L,n}$ and $f_{R,n}$ satisfy the bulk equation of motion
\begin{align}
\left[\partial_5^2+\left(\frac{a^\prime}{a}+2M\right)\partial_5+\frac{m_n^2}{a^2}\right]f_{L,n}&=0,\\
\left[\partial_5^2+\left(\frac{a^\prime}{a}-2M\right)\partial_5+\frac{m_n^2}{a^2}\right]f_{R,n}&=0,
\end{align} 
with the solution
\bea
\tilde{S}_{M_f}(x_5,m_n)=\frac{\pi m_n}{2k}a(x_5)^{\mp c-\frac{1}{2}}\left[J_{\frac{1}{2}\pm c}\left(\frac{m_n}{k}\right)Y_{\frac{1}{2}\pm c}\left(\frac{m_n}{ka(x_5)}\right)-Y_{\frac{1}{2}\pm c}\left(\frac{m_n}{k}\right)J_{\frac{1}{2}\pm c}\left(\frac{m_n}{ka(x_5)}\right)\right]\nn\\
\eea
where $M=-ck$.
In the absence of mixings between components with different B.C., only the components with the B.C. $(++)$ have the zero modes.
For the fields charged under QCD color, the zero modes of the bulk fields $\xi_q$ and $\xi_t$ can be identified as $q_L$ and $t_R$, respectively; for the fields charged under the color in the hidden sector, the zero modes are the vector-like top partners. After those brane localized terms are introduced, different multiplets, such as $\xi_q$ and $\xi_t$, are related via the equation of motion, and induce the Yukawa coupling between the zero modes, such as $q_L$ and $t_R$. These terms are crucial to generate the Higgs potential and trigger EWSB.
On the other hand, these terms make the task of finding the KK spectrum of a generic five-dimensional setup technically quite complicated.
Therefore, instead of pursuing the cumbersome KK decomposition, we resort to the alternative holographic approach to extract the four-dimensional low energy Lagrangian from the five-dimensional setup.

\section{Effective Boundary Action in Holographic Approach}
\label{sec:holography}

The holographic approach~\cite{Maldacena:1997re,ArkaniHamed:2000ds,Rattazzi:2000hs,PerezVictoria:2001pa,Barbieri:2003pr,Contino:2004vy} is extremely useful when computing the low energy observables, such as the Higgs potential, in the five-dimensional gauge-Higgs unification scenario. In the holographic approach, the information of the five-dimensional bulk is encoded in the so-called holographic Lagrangian defined on the UV brane, accordingly with the holographic fields. Equivalently, one can think of the holographic Lagrangian as the low energy effective description of the five-dimensional setup, with the bulk fields being integrated out, while its value on the UV brane being fixed.

More practically, it is much more convenient to work in the gauge where the dependence on $F_5$ is shifted to the boundary action on the UV brane; and it can be achieved by certain gauge transformations with the Wilson line defined in Eq.~\ref{wilsonline}. Accordingly, the boundary conditions need to be redefined as 
\bea
\phi_0\to \Omega \ \phi_0,
\label{UVboundary}
\eea
where $\phi_0\equiv\phi(x_\mu, z=L_0)$ denotes the generic boundary field on the UV brane.
This means when solving the holographic Lagrangian one can set $F_5=0$ temporarily and then use the boundary condition in Eq.~\ref{UVboundary} to recover the Higgs dependence.

Let us solve the holographic Lagrangian for the theory defined in Sec.~\ref{sec:model}. For the purpose of demonstrating the electroweak vacuum can naturally align along the direction with small $\theta\sim v^2/f^2$, we focus on the fermion sector in this section.
Under the field variations $\delta\xi_{q}$ and $\delta\xi_{t}$, the bulk action for the SM sector would vary accordingly as
\begin{align}
\delta S_{fermion}=&\frac{1}{g_5^2}\int \text{d}^4x\int_{L_0}^{L_1}\text{d}z\sqrt{g}\left[\delta\bar{\xi}_{q,t}D\xi_{q,t}+\overline{D\xi}_{q,t}\delta\xi_{q,t}\right]\nn\\
+&\frac{1}{2}\frac{1}{g_5^2}\int\text{d}^4x\sqrt{-g_{\text{ind}}}\left\{\left(\bar{\xi}_{qL}\delta\xi_{qR}+\delta\bar{\xi}_{qR}\xi_{qL}-\bar{\xi}_{qR}\delta\xi_{qL}-\delta\bar{\xi}_{qL}\xi_{qR}\right)|^{L_1}_{L_0}\right.\nn\\
&\ \ \ +\left.\left(\bar{\xi}_{tL}\delta\xi_{tR}+\delta\bar{\xi}_{tR}\xi_{tL}-\bar{\xi}_{tR}\delta\xi_{tL}-\delta\bar{\xi}_{tL}\xi_{tR}\right)|^{L_1}_{L_0}\right\}\ .
\label{vari}
\end{align}
In the above equation, we see that there are non-vanishing boundary terms in $\delta S_{fermion}$.
Only those boundary terms will survive after imposing the bulk equation of motion, namely $D\xi_{q}=0$ and $D\xi_{t}=0$ where $D$ is the five-dimensional Dirac operator. 
The variation of the action vanishes, i.e.,
\bea
\delta S=0.
\label{holovari}
\eea
Therefore, additional boundary terms are needed to cancel the non-vanishing terms in Eq.~\ref{vari}.

It is natural to choose $\xi_{qL}$ and $\xi_{tR}$ as the holographic source fields, as they are the fields with Neumann B.C. according to Eq.~\ref{bulksm}. Then variations of the fields $\xi_{qL}$ and $\xi_{tR}$ vanish on the UV brane as their field value are fixed, 
\bea
\delta\xi_{qL}(z=L_0)=\delta\xi_{tR}(z=L_0)=0.
\label{holofields}
\eea 
On the other hand, $\xi_{qR}(z=L_0)$ and $\xi_{tL}(z=L_0)$ are free to vary. After setting Eq.~\ref{holofields}, we have
\bea
\delta S_{fermion}|_{z=L_0}=\frac{1}{2g_5^2}\int\text{d}^4x\sqrt{-g_{\text{ind}}}\left[-\left(\bar{\xi}_{qL}\delta\xi_{qR}+\delta\bar{\xi}_{qR}\xi_{qL}\right)+\left(\bar{\xi}_{tR}\delta\xi_{tL}+\delta\bar{\xi}_{tL}\xi_{tR}\right)\right]|_{z=L_0}\nn\\
\eea
for the variation of action on the UV brane.
In order to satisfy Eq.~\ref{holovari}, the following terms on the UV brane are introduced
\bea
\Delta S_{UV}=\frac{1}{2}\frac{1}{g_5^2}\int \text{d}^4x\sqrt{-g_{\text{ind}}}\left[(\bar{\xi}_{qL}\xi_{qR}+\bar{\xi}_{qR}\xi_{qL})-(\bar{\xi}_{tR}\xi_{tL}+\bar{\xi}_{tL}\xi_{tR})\right]|_{z=L_0}\ ,
\eea
and then we see that 
\bea
\delta S_{fermion}|_{z=L_0}+\delta(\Delta S_{UV})=0\ .
\eea

As we mentioned, to generate the top Yukawa and hence induce the Higgs potential radiatively, additional mixing term in Eq.~\ref{IR1} on the IR brane needs to be introduced to explicitly break the $SO(5)$ symmetry. Note that both the top quark mass and Higgs potential vanish when $m\rightarrow 0$, which renders the Higgs boson an exact Goldstone boson. With the additional terms on the IR brane
\bea
\Delta S_{IR}=\frac{1}{2}\frac{1}{g_5^2}\int\text{d}^4x\sqrt{-g_{\text{ind}}}\left[-\left(\bar{\xi}_{qL}\xi_{qR}+\bar{\xi}_{qR}\xi_{qL}\right)+\left(\bar{\xi}_{tL}\xi_{tR}+\bar{\xi}_{tR}\xi_{tL}\right)\right]|_{z=L_1}\ ,
\eea
as well as the requirement that the variation on the IR brane vanishes, namely
\begin{align}
\delta S_{fermion}|_{z=L_1}+\delta(\Delta S_{IR})+\delta S_{\text{mixing}}=0
\end{align}
where $S_{\text{mixing}}$ is defined in Eq.~\ref{IR1}.
the boundary conditions on the IR brane can be derived as
\bea
\begin{aligned}
\xi_{tL}(z=L_1)&=-m\ \xi^{(1)}_{qL}(z=L_1),\\
\xi^{(1)}_{qR}(z=L_1)&=m\ \xi_{tR}(z=L_1),\\
\xi^{(4)}_{qR}(z=L_1)&=0.
\end{aligned}
\label{bcsm}
\eea
Turning off the mixings, then the boundary conditions in Eq.~\ref{bcsm} would reduce to $\xi_{tL}(z=L_1)=0$ and $\xi_{qR}(z=L_1)=0$, which is fully consistent with the B.C. assignment in Eq.~\ref{bulksm}. To obtain the holographic Lagrangian, one needs to solve the fields $\xi_{tL}$ and $\xi_{qR}$ in terms of the holographic source fields, with the boundary conditions obtained as in Eq.~\ref{bcsm}.

Due to the mixing term introduced in Eq.~\ref{IR2}, the non-vanishing Higgs potential can also arise from the hidden sector when $\widetilde{m}$ is not zero. 
For the bulk fermions in the hidden sector, it is natural to choose $\xi_{\widetilde{q}L}=(\xi^{(4)}_{\widetilde{q}L}, \ \xi^{(1)}_{\widetilde{q}L})^T$ and $\xi_{\widetilde{T}R}$ as the holographic fields. Similar to the previous situation, they are the fields with the Neumann B.C. on the UV brane. Following the holographic approach, we obtain the IR-boundary conditions for the hidden sector as
\bea
\begin{aligned}
\xi_{\widetilde{T}L}(z=L_1)&=-\widetilde{m}\ \xi^{(1)}_{\widetilde{q}L}(z=L_1),\\
\xi^{(1)}_{\widetilde{q}R}(z=L_1)&=\widetilde{m}\ \xi_{\widetilde{T}R}(z=L_1),\\
\xi^{(4)}_{\widetilde{q}R}(z=L_1)&=0,
\end{aligned}
\label{bchidden}
\eea
which is irrelevant to the extra action defined on the UV brane in Eq.~\ref{mass}. Since the four-dimensional field $\widetilde{q}_{0R}$ is treated as a holographic field, the variation of $S_{mass}$ then automatically vanish.

\section{The effective Lagrangian and Higgs Potential}
\label{sec:potential}
In the last section, we follow the standard procedure to solve the fermion sector of the five-dimensional model holographically~\cite{Contino:2004vy}, and we end up with the effective boundary action, in which the mixings on the IR brane are crucial for breaking the shift symmetry of Nambu-Goldstone Higgs explicitly. In this section, we present the resulting low-energy holographic Lagrangian which only contains the UV fields, and furthermore from that we derive the Higgs potential regarding the fermionic contribution. 

After solving the boundary conditions obtained in Eqs.~\ref{bcsm} and~\ref{bchidden}, we obtain the low energy effective Lagrangian parametrized as the following
\bea
\mathcal{L}_{\text{eff}}=\ &\bar{t}_L p\!\!\!/\Pi_{t_L}t_L+\bar{t}_R p\!\!\!/\Pi_{t_R}t_R-\bar{t}_L\Pi_{t_Lt_R}t_R+\bar{\widetilde{L}} p\!\!\!/\widetilde{\Pi}_{L}\widetilde{L}+\bar{\widetilde{R}} p\!\!\!/\widetilde{\Pi}_{R}\widetilde{R}-\bar{\widetilde{L}}\widetilde{\Pi}_{LR}\widetilde{R}+\textnormal{h.c.}\ ,
\label{holo}
\eea
where, for compactness, the holographic source fields in the color-neutral sector are denoted as
\bea
\widetilde{L}=
\left(
\baa{c}
\widetilde{t}_L\\ \widetilde{T}_L
\eaa
\right),\ \ 
\widetilde{R}=
\left(
\baa{c}
\widetilde{t}_R\\ \widetilde{T}_R
\eaa
\right),\ \ \ 
\eea
and the information of strong dynamics is encoded in the form factors $\Pi_{t_L}, \Pi_{t_R}, \Pi_{t_Lt_R}$ and $\widetilde{\Pi}_{L}, \widetilde{\Pi}_{R}, \widetilde{\Pi}_{LR}$. In the above effective Lagrangian, we neglect the bottom sector as they are less relevant to EWSB due to the smallness of the bottom Yukawa coupling.

The Higgs dependence can be read off after the ``Goldstone matrix" is dressed on. Accordingly, the holographic fields need to be redefined as
\bea
\xi_{qL}&=\frac{1}{\sqrt{2}}
\left(
\baa{c}
b_L\\-ib_L\\t_L\\it_L\\0 
\eaa
\right)\to \frac{1}{\sqrt{2}}
\left(
\baa{ccc}
\bold1_{3\times 3}&\ &\ \\
\ &c_h&s_h\\
\ &-s_h&c_h\\  
\eaa
\right) 
\left(
\baa{c}
b_L\\-ib_L\\t_L\\it_L\\0 
\eaa
\right)=\frac{1}{\sqrt{2}}
\left(
\baa{c}
b_L\\-ib_L\\t_L\\it_Lc_h\\-it_Ls_h
\eaa
\right),
\eea
and
\bea
\xi_{\widetilde{q1}L}&=\frac{1}{\sqrt{2}}
\left(
\baa{c}
\widetilde{b}_L\\-i\widetilde{b}_L\\\widetilde{t}_L\\i\widetilde{t}_L\\\sqrt{2}\widetilde{T}_L 
\eaa
\right)\to\frac{1}{\sqrt{2}}\left(
\baa{ccc}
\bold1_{3\times 3}&\ &\ \\
\ &c_h&s_h\\
\ &-s_h&c_h\\  
\eaa
\right) \left(
\baa{c}
\widetilde{b}_L\\-i\widetilde{b}_L\\\widetilde{t}_L\\i\widetilde{t}_L\\\sqrt{2}\widetilde{T}_L 
\eaa
\right)=
\frac{1}{\sqrt{2}}\left(
\baa{c}
\widetilde{b}_L\\-i\widetilde{b}_L\\\widetilde{t}_L\\i\widetilde{t}_Lc_h+\sqrt{2}\widetilde{T}_Ls_h\\\sqrt{2}\widetilde{T}_Lc_h-i\widetilde{t}_Ls_h
\eaa
\right),
\eea
and
\bea
\widetilde{q}_R&=\frac{1}{\sqrt{2}}
\left(
\baa{c}
\widetilde{b}_R\\-i\widetilde{b}_R\\\widetilde{t}_R\\i\widetilde{t}_R\\0 
\eaa
\right)\to
\frac{1}{\sqrt{2}}
\left(
\baa{ccc}
\bold1_{3\times 3}&\ &\ \\
\ &c_h&s_h\\
\ &-s_h&c_h\\  
\eaa
\right) 
\left(
\baa{c}
\widetilde{b}_R\\-i\widetilde{b}_R\\\widetilde{t}_R\\i\widetilde{t}_R\\0 
\eaa
\right)=\frac{1}{\sqrt{2}}
\left(
\baa{c}
\widetilde{b}_R\\-i\widetilde{b}_R\\\widetilde{t}_R\\i\widetilde{t}_Rc_h\\-i\widetilde{t}_Rs_h
\eaa
\right).
\eea

Here we report the explicit holographic Lagrangian. In the SM top sector, we have
\bea
\mathcal{L}^H_{SM}=&\ \bar{t}_L\Pi_{L}(0)\frac{p\!\!\!/}{p}t_L+\frac{\Pi_{L}(m)-\Pi_{L}(0)}{2}\bar{t}_L\frac{p\!\!\!/}{p}s_h^2t_L+\bar{t}_R\frac{p\!\!\!/}{p}\Pi_{R}(m)t_R+\frac{-i\ \Pi_{LR}(m)}{\sqrt{2}}s_h\bar{t}_Lt_R+\text{h.c.}\ ;
\label{effholosm}
\eea
while in the color-neutral sector, we have
\begin{align}
\mathcal{L}^H_{hidden}&= \bar{\widetilde{t}}_L\frac{p\!\!\!/}{p}\widetilde{\Pi}_L(0)\widetilde{t}_L+\bar{\widetilde{T}}_L\frac{p\!\!\!/}{p}\widetilde{\Pi}_L(\widetilde{m}^\prime)\widetilde{T}_L\nn\\
&+\left(\widetilde{\Pi}_L(\widetilde{m}^\prime)-\widetilde{\Pi}_L(0)\right)\times\left(\frac{1}{2}\bar{\widetilde{t}}_L\frac{p\!\!\!/}{p}s_h^2\widetilde{t}_L-\bar{\widetilde{T}}_L\frac{p\!\!\!/}{p}s_h^2\widetilde{T}_L+\frac{i}{\sqrt{2}}\bar{\widetilde{t}}_L\frac{p\!\!\!/}{p}s_hc_h\widetilde{T}_L\right)\nn\\
&+\bar{\widetilde{t}}_R p\!\!\!/ \widetilde{t}_R+\bar{\widetilde{T}}_R\frac{p\!\!\!/}{p}\widetilde{\Pi}_R(\widetilde{m}^\prime)\widetilde{T}_R-\widetilde{m}_{q}\bar{\widetilde{t}}_L\widetilde{t}_R+\widetilde{\Pi}_{LR}(\widetilde{m}^\prime)\left(\frac{i}{\sqrt{2}}\bar{\widetilde{t}}_L\ s_h+\bar{\widetilde{T}}_L\ c_h\right)\widetilde{T}_R+\text{h.c.}\ .
\label{effholohidden}
\end{align}
In these form factors, the Higgs dependence can also be inferred according to the quantum numbers of the holographic fields under $SU(2)_L$ group. 
For example, $\Pi_{t_Lt_R}$ should be proportional to $s_h\equiv \sin(h/f)$ since $t_L$ belongs to the $SU(2)_L$ doublet while $t_R$ is a singlet. On the other hand, the wave functions should always contains Higgs dependence like $s_h^2$ or $c_h\equiv \cos(h/f)$.
 Furthermore, we see all the Higgs dependence in the holographic Lagrangian in Eq.~\ref{effholosm} and Eq.~\ref{effholohidden} would vanish if the mixing terms on the IR brane were turned off, i.e., $m, \widetilde{m}^\prime\to 0$. Note that both $\Pi_{LR}(m)$ and $\widetilde{\Pi}_{LR}(\widetilde{m}^\prime)$ are proportional to $m$ and $\widetilde{m}^\prime$, respectively. We also note that the mass term of the doublet defined in Eq.~\ref{mass} is invariant after dressing the Higgs field.

\begin{figure}
\includegraphics[scale=0.55]{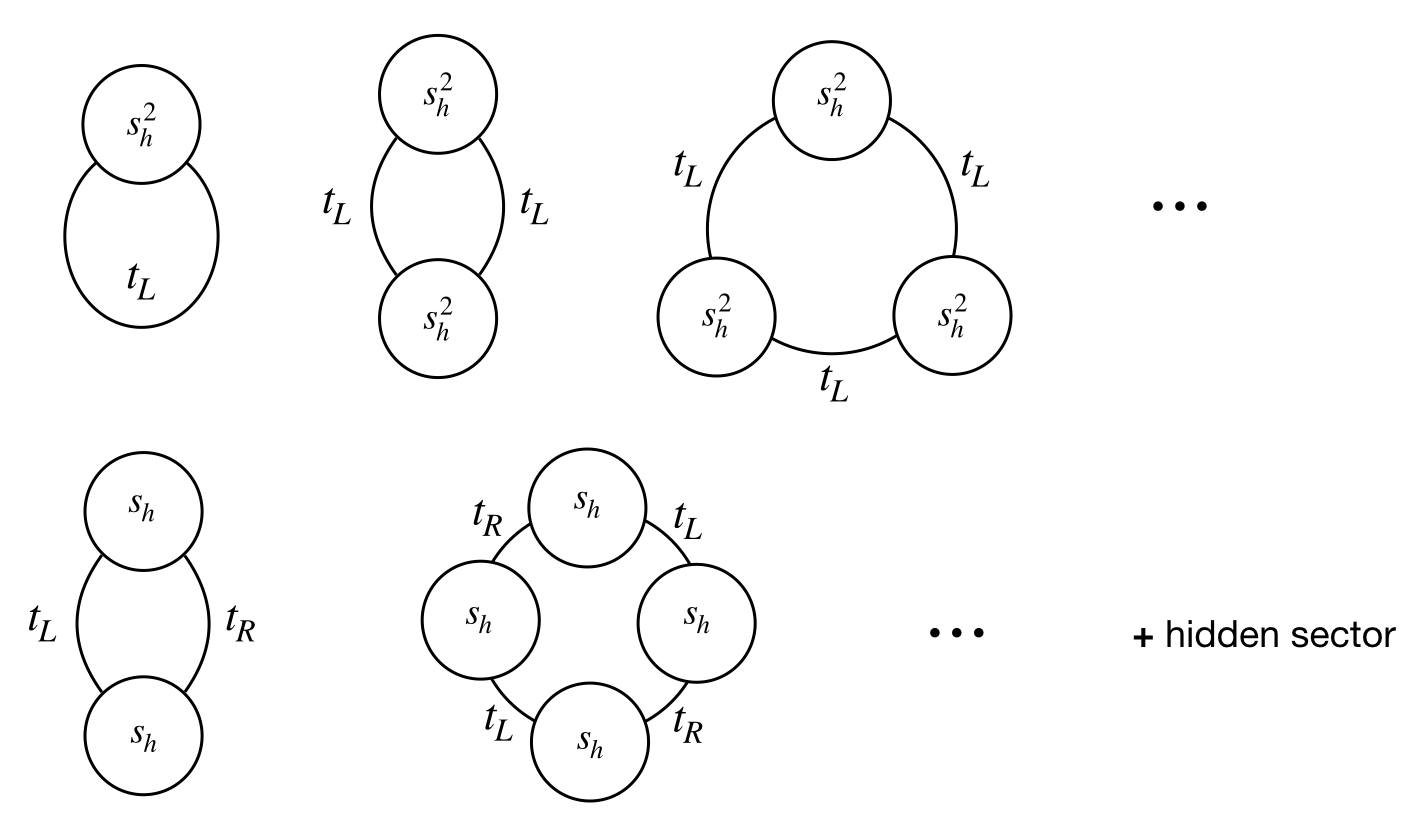}
\caption{The Feynman diagrams for the holographic model. In each blob we show the Higgs dependence for the corresponding vertex, while with each line connecting blobs we show the corresponding holographic field propagating in the loop. We only show the leading Feynman diagrams in the SM sector with and without chirality flip. Despite of the fact that the Higgs dependence is more complicated, the same procedure can be followed when obtaining the Feynman diagrams in the hidden sector. The compact form as in Eq.~\ref{holo} is useful to systematically lay out the relevant Feynman diagrams. The full form of the Higgs potential in Eq.~\ref{potential_composite} is obtained after summing over all the diagrams with appropriate symmetry factors being taken into account. }
\label{feyn_holo}
\end{figure}

Based on the above holographic Lagrangian in Eq.~\ref{effholosm} and Eq.~\ref{effholohidden}, we obtain the Higgs potential, including both the contribution of the SM sector and the color-neutral sector, as
\begin{align}
&V(h)=-\frac{2N_c}{16\pi^2}\int \textnormal{d}Q^2 Q^2\ \textnormal{log}\left[\Pi_{t_L}\Pi_{t_R}\cdot Q^2+|\Pi_{t_Lt_R}|^2\right]\nn\\
&-\frac{2\widetilde{N}_c}{16\pi^2}\int \textnormal{d}Q^2 Q^2\ \textnormal{Tr}\left\{\textnormal{log}\left(Q^2+\widetilde{\Pi}_{LR}\widetilde{\Pi}^{-1}_{R}\widetilde{\Pi}^\dagger_{LR}\widetilde{\Pi}^{-1}_{L}\right)+\textnormal{log}\left(1+(\widetilde{\Pi}_{L}-\widetilde{\Pi}_{L0})\widetilde{\Pi}^{-1}_{L0}\right)\right.\nn\\
&\quad\quad\quad\quad\quad\quad\quad\quad\quad\quad\quad\quad+\left.\textnormal{log}\left(1+(\widetilde{\Pi}_{R}-\widetilde{\Pi}_{R0})\widetilde{\Pi}^{-1}_{R0}\right)\right\},
\label{potential_composite}
\end{align}
where the loop momentum is rotated into the Euclidean space, i.e., $Q^2=-p^2$, and $\widetilde{\Pi}_{L0, R0}$ denote the Higgs-independent part of the wave functions defined in Eq.~\ref{holo}. According to the Higgs dependence, the potential can be reorganized into 
\begin{align}
V(h)=-\frac{2N_c}{16\pi^2}&\int \textnormal{d}Q^2 Q^2\ \left\{\frac{\Pi_{t_L}-\Pi_{t_L0}}{\Pi_{t_L0}} -\frac{1}{2} \left(\frac{\Pi_{t_L}-\Pi_{t_L0}}{\Pi_{t_L0}}\right)^2+ \frac{1}{3} \left(\frac{\Pi_{t_L}-\Pi_{t_L0}}{\Pi_{t_L0}}\right)^3+\cdots\right.\nn\\
&\ \ \ +\left. \frac{|\Pi_{t_Lt_R}|^2}{\Pi_{t_L}\Pi_{t_R}\cdot Q^2} -\frac{1}{2} \left(\frac{|\Pi_{t_Lt_R}|^2}{\Pi_{t_L}\Pi_{t_R}\cdot Q^2}\right)^2+ \frac{1}{3} \left(\frac{|\Pi_{t_Lt_R}|^2}{\Pi_{t_L}\Pi_{t_R}\cdot Q^2}\right)^3+\cdots\right\}\nn\\
&+\ hidden\ sector
\end{align}
where $\Pi_{t_L0}$ denotes the Higgs-independent part of $\Pi_{t_L}$. We demonstrate the series of the Higgs potential terms using Feynman diagrams as shown in Fig.~\ref{feyn_holo}.
 Diagrams in the hidden sector can be laid out in a similar way, with corresponding form factors being organized in the matrix form as in Eq.~\ref{holo} accordingly.
As the gauge sector tends to preserve the EW vacuum, we omit its contribution in this section. Since our purpose here is to show vacuum misalignment can be realized even considering only the fermionic contribution.

\begin{figure}
\includegraphics[scale=0.35]{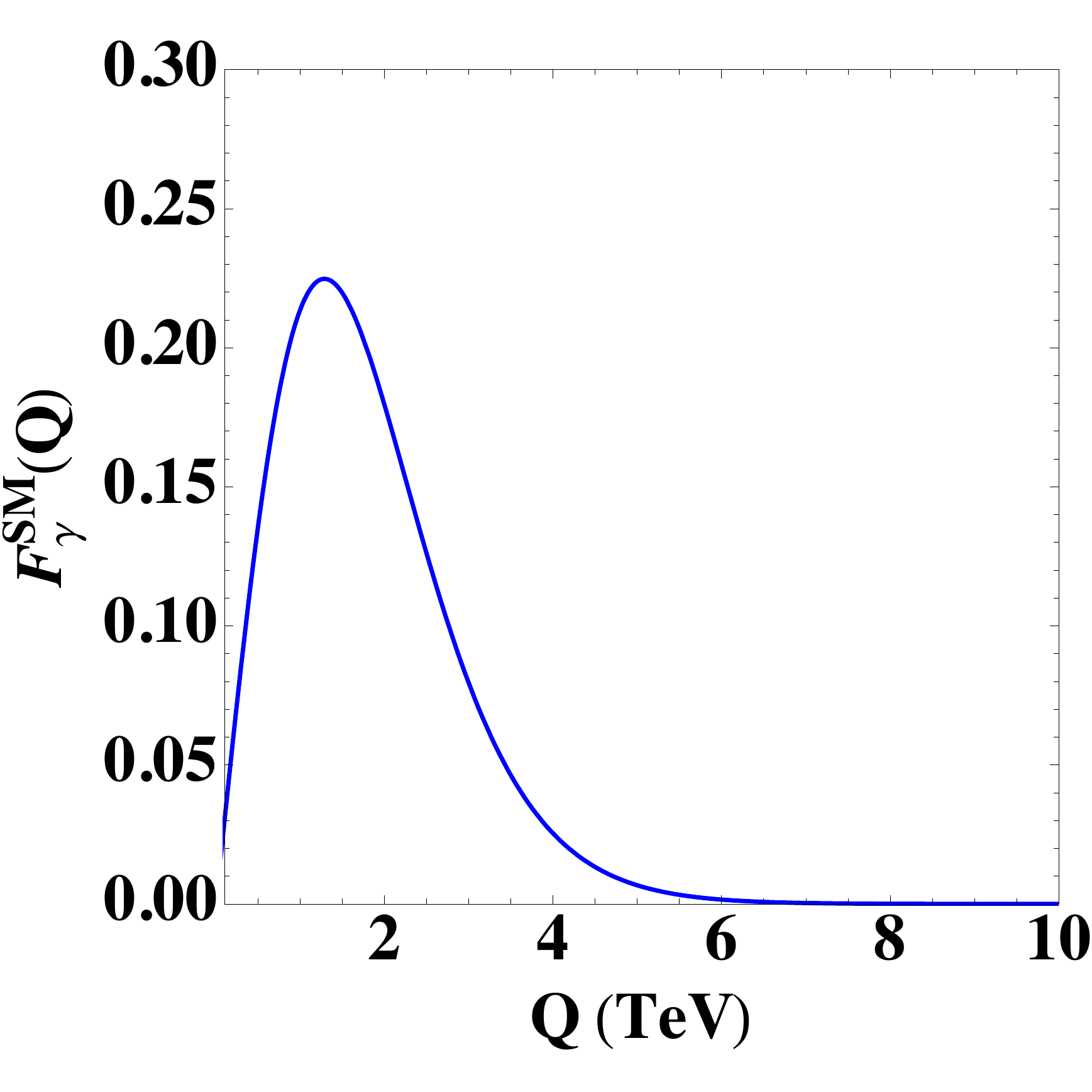}
\includegraphics[scale=0.35]{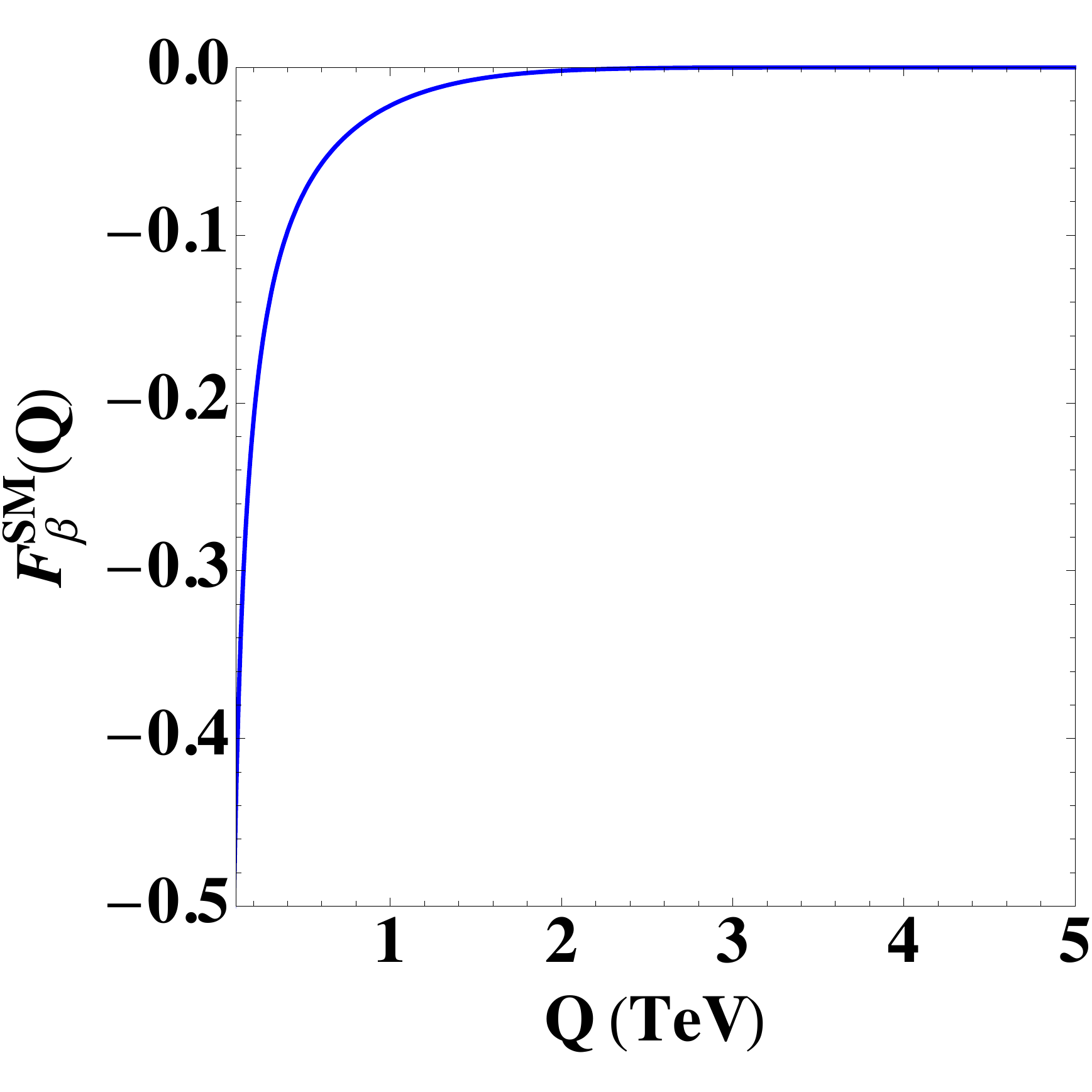}
\includegraphics[scale=0.35]{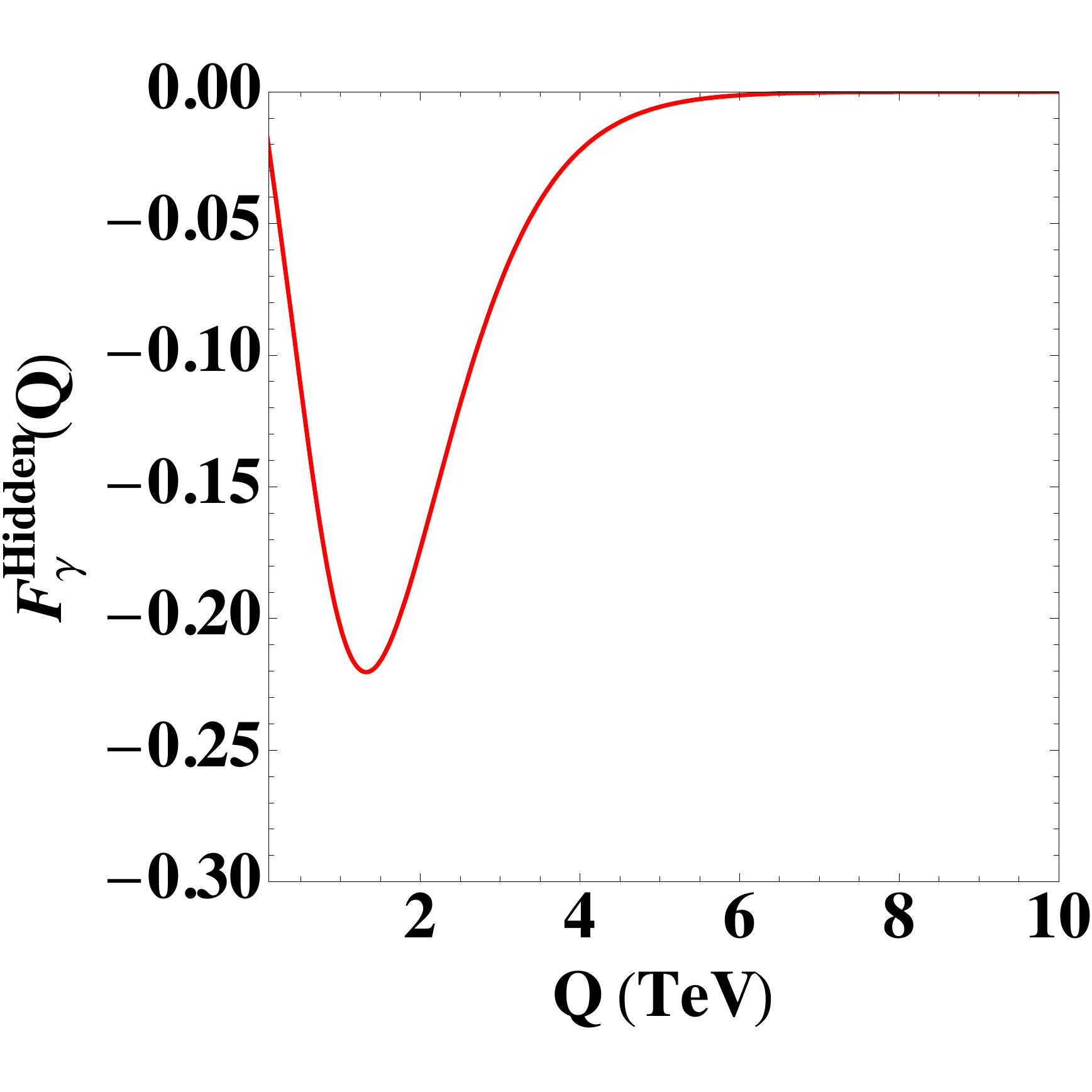}
\includegraphics[scale=0.35]{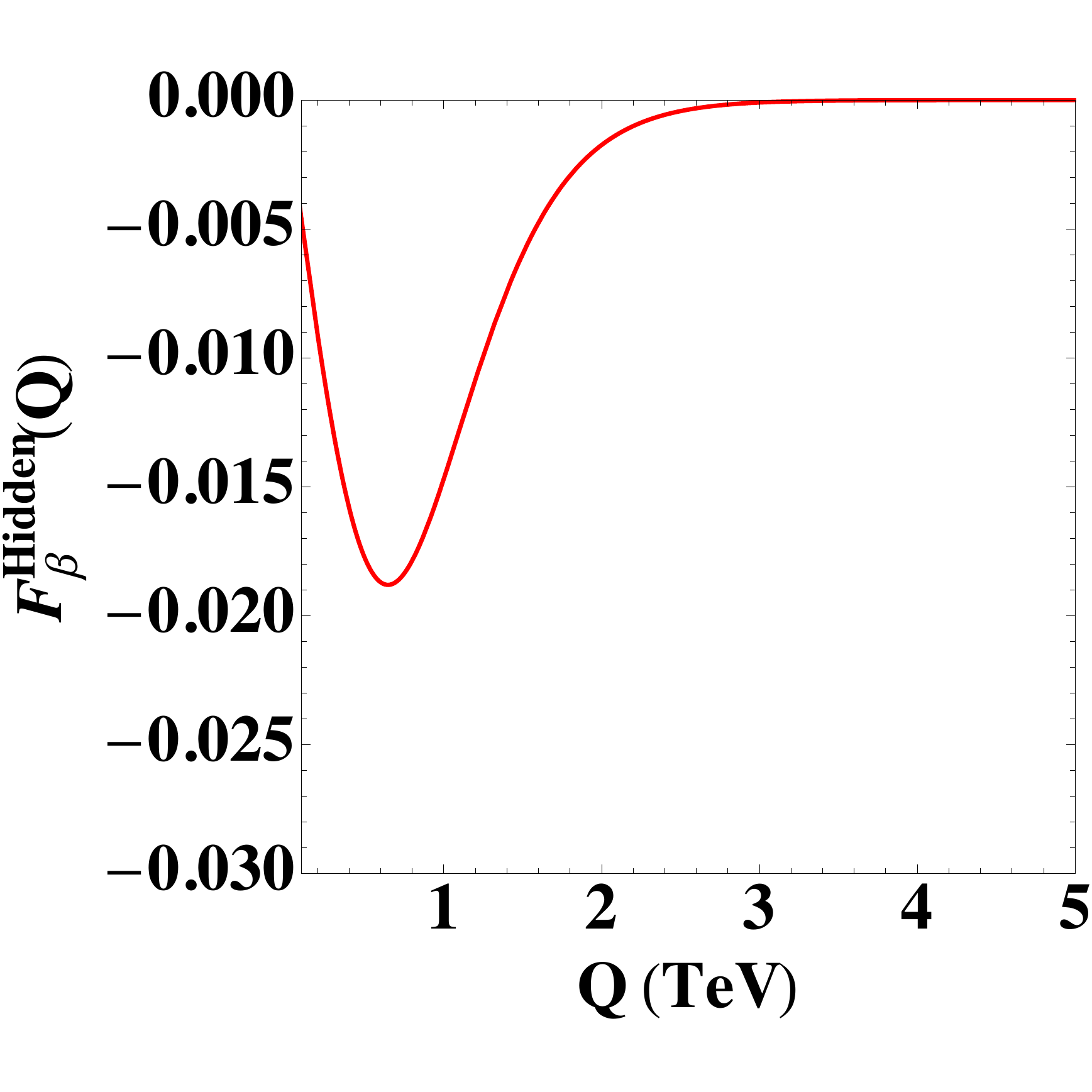}
\caption{Illustration of the form factor combinations in the Higgs potential versus the loop momentum $Q$. The blue lines denote the contribution from the SM sector to $\gamma_f$ (top left) and $\beta_f$ (top right), while the red lines denote the contribution from the color neutral sector to $\gamma_f$ (bottom left) and $\beta_f$ (bottom right). At the region $Q\to\infty$, the form factor combinations vanish and thus leads to the finiteness of the Higgs potential.}
\label{composite_FF}
\end{figure}

According to Eq.~\ref{potential_composite}, the general form of the simplified Higgs potential can be parameterized by
\bea
V(h)\simeq &-\gamma_f s_h^2+\beta_f s_h^4+\cdots
\label{potential_simplify}
\eea
with the definitions 
 \begin{align}
 \gamma_f&=\frac{3}{4 \pi^2}\int \text{d}Q \ (F^{\text{SM}}_\gamma+F^{\text{Hidden}}_\gamma)\ ,\nn\\
 \beta_f&=-\frac{3}{4 \pi^2}\int \text{d}Q \ (F^{\text{SM}}_\beta+F^{\text{Hidden}}_\beta)\ ,
 \label{integrands}
 \end{align}
 where higher order terms with the Higgs dependence of $\mathcal{O}(s_h^6)$ are neglected. In the above equation, the SM contribution is 
 \begin{align}
 F^{\text{SM}}_\gamma&=Q^3\left(\frac{\Pi_{t_L}-\Pi_{t_L0}}{\Pi_{t_L0}}+\frac{|\Pi_{t_Lt_R}|^2}{\Pi_{t_L0}\Pi_{t_R}\cdot Q^2}\right)\ ,\nn\\
 F^{\text{SM}}_\beta&=Q^3\left[-\frac{1}{2} \left(\frac{\Pi_{t_L}-\Pi_{t_L0}}{\Pi_{t_L0}}\right)^2-\frac{1}{2} \left(\frac{|\Pi_{t_Lt_R}|^2}{\Pi_{t_L0}\Pi_{t_R}\cdot Q^2}\right)^2-\frac{\Pi_{t_L}-\Pi_{t_L0}}{\Pi_{t_L0}} \cdot \frac{|\Pi_{t_Lt_R}|^2}{\Pi_{t_L0}\Pi_{t_R}\cdot Q^2}\right].
 \end{align}
 We can also have the contribution from the color-neutral sector similarly, although the explicit expressions are much complicated due to different Higgs dependence.
We plot the contribution from the SM sector and the color-neutral sector to the Higgs potential in Fig.~\ref{composite_FF}. 
We see explicitly that the correct vacuum misalignment is realized because the SM sector and the hidden sector cancel each other in $\gamma_f$, and the contribution from the color-neutral sector to $\beta_f$ is much suppressed compared to the SM sector. Note that all the integrands depicted in Fig.~\ref{composite_FF} vanish when $Q\to \infty$ due to the exponential suppression on the dependence of the loop momentum. Thus the Higgs potential is finite, i.e., it does not depend on the new physics scales at far UV. The calculability of the Higgs potential is the general feature in the warped extra-dimensional model, this is the due to the collective symmetry breaking. 
When calculating the coefficients $\gamma_f$ and $ \beta_f$, only the leading terms in Eq.~\ref{potential_composite} are included, this would lead to spurious IR divergence as usual. We therefore choose the IR cutoff as $100$ GeV in this work.
As composite top partners will contribute to the Higgs potential and set the overall scale for the Higgs potential if they contribute individually, the corresponding fine-tuning level of our model is similar to other composite Higgs models~\cite{Panico:2012uw}.

\section{Electroweak Precision Tests}
\label{sec:precision}
Electroweak precision observables~\cite{Peskin:1991sw,Barbieri:2004qk} are potentially very sensitive to new physics, and therefore can be used to set a lower bound on the new physics scale. 
Let's first consider the effective Lagrangian in the gauge sector which parametrizes the new physics effect from strong dynamics.
In the Landau gauge, the holographic Lagrangian of gauge sector is
\begin{align}
&\mathcal{L}^{H}_{gauge}=\frac{P_t^{\mu\nu}}{2}\left[\frac{2}{g_5^2}W^+_\mu W^-_\nu\left(\Pi^++\frac{s_h^2}{2}\left(\Pi^--\Pi^+\right)\right)+A_\mu A_\nu\left(\frac{2s^2_W}{g_5^2}\Pi^++\frac{c^2_W-s^2_W}{g^2_{X}}\Pi^+_X\right)\right.\nn\\
&\left.+Z_\mu Z_\nu \left(\frac{c^2_W+s^2_Xs^2_W}{g_5^2}\Pi^+ +\frac{c^2_X s^2_W}{g^2_{X}}\Pi^+_X +\frac{s^2_h}{2c^2_Wg^2_5} (\Pi^- -\Pi^+)\right)+Z_\mu A_\nu 2c_Ws_W \left(\frac{c^2_X}{g^2_5}\Pi^+ -\frac{c^2_X}{g^2_{X}}\Pi^+_X\right)\right]
\end{align}
with the four dimensional gauge coupling of the SM defined as
\bea
\frac{\Pi^{+\prime}}{g_5^2}=\frac{\Pi^{+\prime}_X}{g^2_{X}}=\frac{1}{g^2},
\eea
where $\Pi^\prime=\partial \Pi(p^2)/\partial p^2|_{p^2=0}$.
With the above holographic Lagrangian, one can straightforwardly calculate the leading contribution to the so-called oblique parameters
\begin{align}
\frac{\alpha}{4 s_W^2 c_W^2} S&=\Pi^\prime_{ZZ}-\frac{c_W^2-s_W^2}{c_Ws_W}\Pi^\prime_{Z\gamma}-\Pi^\prime_{\gamma\gamma},\\
\alpha T&=\frac{\Pi_{WW}(0)}{m_W^2}-\frac{\Pi_{ZZ}(0)}{m_Z^2}.
\end{align}
The leading order contribution is found to be
\begin{align}
S&=\frac{8\pi s_h^2}{g_5^2}\left(\Pi^{+\prime}-\Pi^{-\prime}\right)\simeq \frac{3}{2}\frac{\pi v^2}{M^2_{KK}}\simeq \frac{3}{2} \pi v^2 L_1^2,\label{leadingS}\\
T&=0\quad \quad \text{at\ tree\ level}\ .
\end{align}
Note that the $T$ parameter vanishes at the leading order because of the custodial symmetry. However, the non-vanishing $S$ parameter leads to constraint on the mass of the lightest spin-$1$ resonance. Given the fixed value of $v$, the correction from strong dynamics to $S$ can be sufficiently suppressed if all the KK resonances decouple as $M^2_{KK}\to\infty$.

Beyond the leading order, there are two kinds of contributions to the $T$ parameter: UV contribution from the composite states, and IR contribution from the Higgs loop.

First let us consider the loop corrections from the strong dynamics. The nonzero corrections to $T$ parameter will in general arise from the loops of heavy fermions and vectors, e.g. the bi-doublet fermions under $SU(2)_L\times SU(2)_R$ with the custodial-symmetry-violating boundary conditions~\cite{Carena:2006bn, Carena:2007ua, Anastasiou:2009rv}. With the nature of strong dynamics, the size of these contributions are estimated to be of the order as~\cite{Giudice:2007fh,Contino:2017moj}
\bea
\Delta T^{_{\rm Loop}}_{_{\rm UV}}\sim \frac{N_c}{16 \pi^2}\frac{(y_L^4+\widetilde{y}_L^4) f}{M_{KK}} \xi,
\eea
where $y_L$ and $\widetilde{y}_L$ denote the mixing parameters.
Due to symmetry protection, the $T$ parameter is usually calculable in this case. For the contribution from the zero modes of the color-neutral sector, one can see their contribution is sufficiently small due to the relative small mass splitting inside the doublet~\cite{Xu:2018ofw}.

At low energies, due to Higgs nonlinearity ($\xi\neq 0$), the couplings of the composite Higgs and electroweak gauge bosons are universally modified. Calculated at one-loop order for the oblique parameters, the corrections from the Higgs loop read 
\begin{align}
\Delta S^{_{\rm Loop}}_{_{\rm IR}}=&+ \frac{\xi}{12\pi}\ \text{log}\left(\frac{\Lambda^2}{m_h^2}\right),\label{nonlinearS}\\
\Delta T^{_{\rm Loop}}_{_{\rm IR}}=&- \frac{3\xi}{16 \pi}\frac{1}{\cos^2\theta_W}\ \text{log}\left(\frac{\Lambda^2}{m_h^2}\right)\label{nonlinearT},
\end{align}
where the cutoff scale $\Lambda\sim M_{KK}$ is usually regulated by the mass scale of composite resonances.

The total contributions to the oblique parameters are the sum of these contributions. Typically  there are compensations between different contributions to the oblique parameters $S$ and $T$~\cite{Carena:2006bn, Carena:2007ua, Anastasiou:2009rv}. Depending on five dimensional parameters, i.e., the bulk masses and mixings in the IR, the positive contributions to the $T$ parameter from the fermion loop is viable~\cite{Carena:2006bn, Carena:2007ua, Anastasiou:2009rv}, which can be used to cancel the negative contribution from Higgs nonlinearity $\Delta T^{_{\rm Loop}}_{_{\rm IR}}$. Thus the bound on $\xi$ and $M_{KK}$ is likely alleviated. Finally we would like to mention that the parameter $\xi$ is also constrained by the Higgs coupling measurements.

\section{Deconstructed Four-Dimensional Model}
\label{sec:deconstruction}
The above five-dimensional setup can transparently be translated into four-dimensional composite Higgs models following the general idea of dimensional deconstruction~\cite{ArkaniHamed:2001ca, Contino:2006nn, Cheng:2006ht}. Under the paradigm of partial compositeness~\cite{Kaplan:1991dc}, there are the external elementary sector, the composite sector, and the mixing sector connecting the previous two and at the same time explicitly breaking the global symmetry. 
\bea
\mathcal{L}=\mathcal{L}_{elem.}+\mathcal{L}_{comp.}+\mathcal{L}_{mix.}
\eea
Because of the large top quark mass, the top quark is believed as a mixed state between the composite sector and the elementary sector. As usual, the top sector and its color-neutral counterpart are responsible for triggering EWSB in composite Higgs models. In this section, we will present the deconstructed $2$-site model as a benchmark composite extension of MNNM.

\begin{figure}[!h]
\includegraphics[scale=0.35]{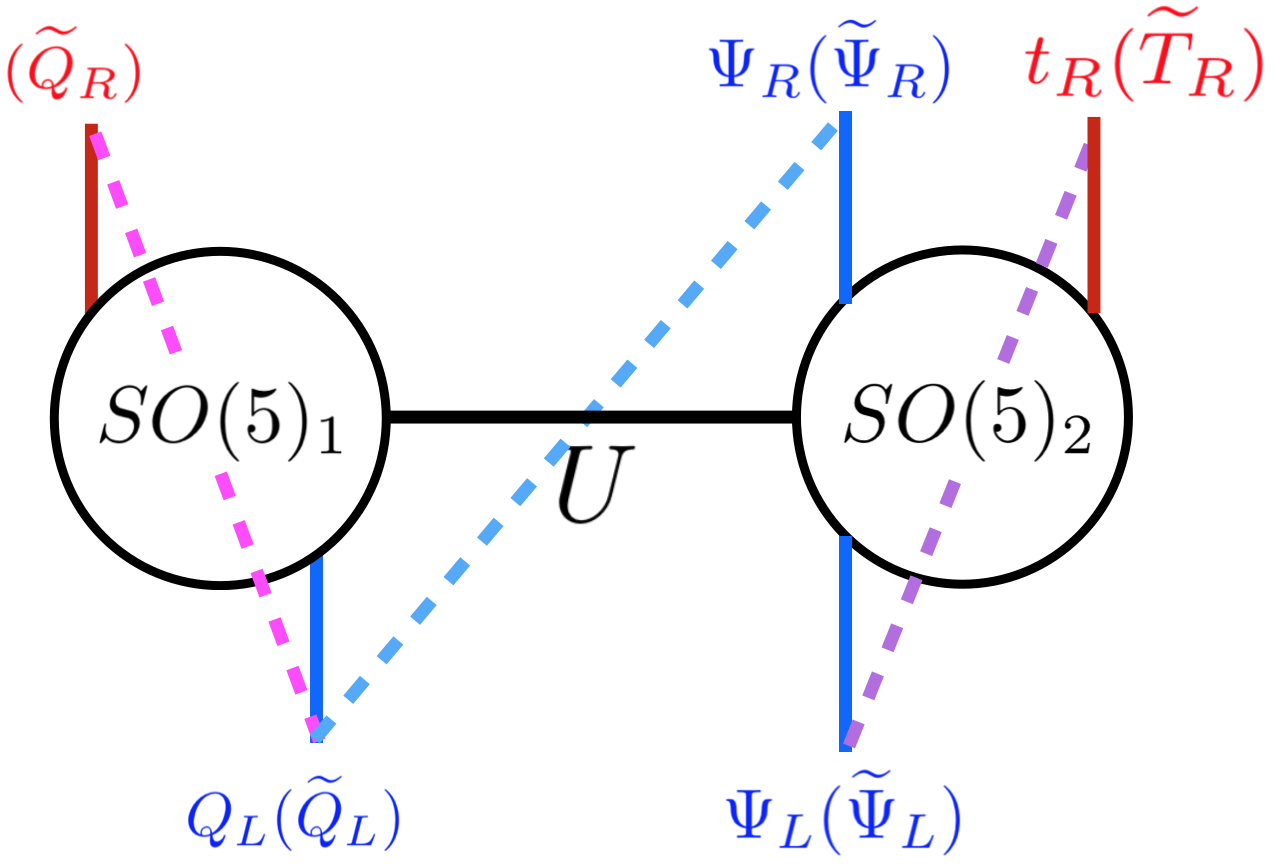}
\caption{The moose diagram for the $2$-site composite realization of minimal neutral naturalness model. Each circle denotes the global symmetry on the site, while the PNGB Higgs is denoted by the link $U$ in between. Accordingly, fermions are embedded into multiplets charged under the global symmetry on each site.}
\label{fig4}
\end{figure}

Based on deconstruction, the simplest one is the $2$-site model based on the global symmetry breaking pattern $SO(5)_1\times SO(5)_2/SO(5)_V$~\cite{Foadi:2010bu,Panico:2011pw} in which the minimal coset $SO(5)/SO(4)$ can be embedded; see the moose diagram in Fig.~\ref{fig4}. 
Considering the bosonic sector, the SM gauge symmetry is identified as the subgroup of $SO(5)_1$, while the $SO(4)$ subgroup of $SO(5)_2$ is identified as the gauge group of composite $\rho$ mesons. Then $6$ of the $10$ Goldstone bosons will be eaten by the $\rho$ mesons after global symmetry breaking, and there will be mixings between the composite mesons and external elementary gauge bosons of the SM. Note that the gauge bosons in the $2$-site model will generally lead to log-divergent Higgs potential, while $3$-site model or multi-site models can render the Higgs potential finite.

At low energies, one can integrate out the composite resonances. Only the SM doublet $q_L=(t_L, b_L)^T$ and singlet $t_R$ and the vector-like doublet $\widetilde{q}_{L,R}$ and singlet $\widetilde{T}_{L,R}$ are treated as dynamical fields. The fields $q_L, \widetilde{q}_L, \widetilde{T}_L$ and $\widetilde{q}_R$ are embedded into multiplets of $SO(5)$, i.e., 
\bea
\begin{aligned}
(Q_L)^I&=(\Lambda_L)^I_{\ i}\ (q_L)^i\ ,\\
(\widetilde{Q}_L)^I&=(\widetilde{\Lambda}_L)^I_{\ i}\ (\widetilde{q}_L)^i+(\widetilde{\Lambda}^\prime_L)^I\ \widetilde{T}_L\ ,\\
(\widetilde{Q}_R)^I&=(\widetilde{\Lambda}_R)^I_{\ i}\ (\widetilde{q}_R)^i,
\end{aligned}
\eea
with ``spurions'' defined as
\bea
\begin{aligned}
\Lambda_L=\widetilde{\Lambda}_L=\widetilde{\Lambda}_R&=\frac{1}{\sqrt{2}}\left(
\baa{ccccc}
0&0&1&i&0\\
1&-i&0&0&0
\eaa
\right)^T\ ,\\
\widetilde{\Lambda}^\prime_L&=\left(
\baa{ccccc}
0&0&0&0&1
\eaa
\right)^T\ .
\end{aligned}
\eea
The fields $t_R$ and $\widetilde{T}_{R}$, on the other hand, are assumed as $SO(5)$ singlets. Explicitly, the $SO(5)$ multiplets are
\bea
Q_L&=\frac{1}{\sqrt{2}}
\left(
\baa{c}
b_L\\-ib_L\\t_L\\it_L\\0 
\eaa
\right),\ \ \ 
\widetilde{Q}_L=\frac{1}{\sqrt{2}}
\left(
\baa{c}
\widetilde{b}_L\\-i\widetilde{b}_L\\\widetilde{t}_L\\i\widetilde{t}_L\\\sqrt{2}\widetilde{T}_L 
\eaa
\right),\ \ \ 
\widetilde{Q}_R&=\frac{1}{\sqrt{2}}
\left(
\baa{c}
\widetilde{b}_R\\-i\widetilde{b}_R\\\widetilde{t}_R\\i\widetilde{t}_R\\0 
\eaa
\right)\subset {\bf{5}}\ ;\label{eq:deconSO5}
\eea
\bea
t_R,\ \ \ \widetilde{T}_R \subset {\bf{1}}.
\eea
We see the assignment of $U(1)_X$ charge is arbitrary for color-neutral top partners, while it is fixed for the SM top quark. Note that $T^3_R=-1/2$ for the doublet $\widetilde{q}_{L,R}$ and $T^3_R=0$ for the singlet $\widetilde{T}_{L,R}$. For completeness, the quantum numbers are summarized in Table~\ref{tab1}.
\begin{table}[!hbp]
\begin{tabular}{r|c|c|c|c}
\hline\hline
\ \ & $q_L$ & $t_R$ & $\widetilde{q}_{L,R}$ & $\widetilde{T}_{L,R}$  \\
\hline
$SU(3)_c$ & $3$ & $3$ & $1$   & $1$  \\
\hline
$SU(3)^\prime_c$ & $1$ & $1$ & $3$   & $3$  \\
\hline
$SU(2)_L$ & $2$ & $1$   & $2$   & $1$\\
\hline
$U(1)_Y$ & $\frac{2}{3}$ & $\frac{2}{3}$ & $-\frac{1}{2} + X$ & $X$ \\
\hline\hline
\end{tabular}
\caption{Quantum numbers for the SM top quark and the elementary color-neutral top partners in the hidden sector. Note the hyper-charge is obtained as $Y=X+T^3_R$.}
\label{tab1}
\end{table}

Considering the fermionic setup, we furthermore introduce composite fermions $\Psi_{L,R}$ and their color-neutral counterparts $\widetilde{\Psi}_{L,R}$, to mimic the KK states of the holographic model. For our purpose, we focus on the Lagrangian which is relevant to generate the Higgs potential and neglect the usual kinetic terms, in which one needs to replace the usual gauge fields with the CCWZ $E_\mu$ for the composite fermions. Given the fermion setup depicted in Fig.~\ref{fig4}, we have the following Lagrangian relevant for studying EWSB,
\bea
\begin{aligned}
\mathcal{L}_{mix.}=\ &yf\bar{Q}_LU\Psi_R-M\bar{\Psi}_{L}\Psi_{R}-m_1\bar{\Psi}_{1L}t_R\\
&+\widetilde{y}f\bar{\widetilde{Q}}_LU\widetilde{\Psi}_R-\widetilde{M}\bar{\widetilde{\Psi}}_{L}\widetilde{\Psi}_{R}-\widetilde{m}_1\bar{\widetilde{\Psi}}_{1L}\widetilde{T}_R-\widetilde{m}_{UV}\bar{\widetilde{Q}}_{L}\widetilde{Q}_{R}+\text{h.c.}\ ,\\
\end{aligned}
\label{composite}
\eea
where $U$ is the Goldstone matrix with its explicit form in unitary gauge as
\bea
U=\left(
\baa{ccc}
\bold1_{3\times 3}&\ &\ \\
\ &c_h&s_h\\
\ &-s_h&c_h\\  
\eaa
\right),
\eea
which corresponds to the Wilson line along the fifth dimension in the five-dimensional model.
Under the gauged $SO(4)$ symmetry, composite $\Psi$ ($\widetilde{\Psi}$) are decomposed into $4$-plet $\Psi_4$ ($\widetilde{\Psi}_4$) and singlet $\Psi_1$ ($\widetilde{\Psi}_1$). Mass splittings of $\Psi_{4,1}$ ($\widetilde{\Psi}_{4,1}$) are assumed to be zero. Because of that, we will see that the Higgs potential is finite even in the deconstructed two-site model. It is interesting to note that the setup in Ref.~\cite{Csaki:2018zzf} is quite similar to ours depicted above.

Due to collective symmetry breaking, the PNGB Higgs is protected by both $SO(5)_1$ and $SO(5)_2$ symmetries. Hence the non-vanishing Higgs potential can only exist when both $SO(5)_1$ and $SO(5)_2$ are explicitly broken. The global symmetry $SO(5)_1$ is explicitly broken in the fermion sector, as fermions forms incomplete multiplets $Q_L$ and $\widetilde{Q}_{L,R}$. On the other hand, the symmetry $SO(5)_2$ is explicitly broken because of the soft parameters $m_1$ and $\widetilde{m}_1$.

Based on the fermionic sector in Eq.~\ref{composite}, we can integrate out the composite states and map it to the effective Lagrangian in Eq.~\ref{holo} via
\bea
\mathcal{L}_{\text{eff}}=-\sum_i\frac{\kappa^\dagger(p\!\!\!/+M_{\Psi_i})\kappa}{p^2-M^2_{\Psi_i}},
\eea
where $\kappa$ denotes the elementary fermions with appropriate Goldstone matrix being dressed on. The explicit form of $\kappa$ depends on fermion representations, e.g., $\kappa$ is $\kappa=U\mathcal{F}$ for the fundamental representation $\mathcal{F}$ of $SO(5)$. Here $\mathcal{F}$ denotes the   $SO(5)$ multiplets defined in Eq.~\ref{eq:deconSO5}. 
Based on the above formula, the explicit form factors can be obtained as
\begin{align}
&\Pi_{t_L}=1-\frac{y^2f^2}{2}\frac{1}{p^2-M^2},\ \ \Pi_{t_R}=1-\frac{m^2_1}{p^2-M^2},\nn\\
&\Pi_{t_Lt_R}=\frac{iyf}{\sqrt{2}}s_h\frac{Mm_1}{p^2-M^2}, 
\label{ffsm}
\end{align}
and
\begin{align}
&\widetilde{\Pi}_L=
\left(
\baa{cc}
1-\frac{\widetilde{y}^2f^2}{2(p^2-\widetilde{M}^2)}& 0\\ 
0 & 1-\frac{\widetilde{y}^2f^2}{p^2-\widetilde{M}^2}
\eaa
\right),\ \nn\\
&\widetilde{\Pi}_R=
\left(
\baa{cc}
1& 0\\ 
0 & 1-\frac{\widetilde{m}_1^2}{p^2-\widetilde{M}^2}
\eaa
\right),\ \nn\\
&\widetilde{\Pi}_{LR}=
\left(
\baa{cc}
\widetilde{m}_{UV}& \frac{-i\widetilde{y}f}{\sqrt{2}}s_h\frac{\widetilde{m}_1\widetilde{M}}{p^2-\widetilde{M}^2}\\ 
0 & \widetilde{y}fc_h\frac{\widetilde{m}_1\widetilde{M}}{p^2-\widetilde{M}^2}
\eaa
\right).
\label{ff}
\end{align}
for the SM sector and the color-neutral sector, respectively.
With specific combination of the form factors, the top quark mass is
\bea
m_t=\frac{\Pi_{t_Lt_R}(p\to 0)}{\sqrt{\Pi_{t_L}(p\to 0)\cdot\Pi_{t_R}(p\to 0)}}\sim y f \langle s_h\rangle\sim y v,
\eea
which is roughly the correct magnitude of the SM value. Similarly, the mass matrix of the color-neutral sectors can be obtained with the momentum $p^2\to 0$. Comparing Eq.~\ref{ff} and Eq.~\ref{effholohidden}, we see the mass matrix of composite model is identical to the one of the holographic model. Both of them can reproduce the low energy spectrum of MNNM, despite some corrections arising from the wave functions.

It is straightforward to see the finiteness of the Higgs potential induced by the form factors in Eq.~\ref{ffsm} and Eq.~\ref{ff}; as the momentum dependence of the relevant form factor combination are
\bea
\lim_{Q^2\to \infty}\frac{|\Pi_{t_Lt_R}|^2}{\Pi_{t_L}\Pi_{t_R}\cdot Q^2}\sim\frac{1}{Q^6}\ ,\ \ \ \lim_{Q^2\to \infty} \frac{1}{Q^2}\rm{Tr}\left[\widetilde{\Pi}_{LR}\widetilde{\Pi}^{-1}_{R}\widetilde{\Pi}^\dagger_{LR}\widetilde{\Pi}^{-1}_{L}\right]\sim\frac{1}{Q^6}\ .
\eea
Beyond two-site, we can introduce more layers of composite states and organize them in a manner that the finite Higgs potential is generally obtained, namely $n$-site models. Based on the naive dimensional analysis (NDA) power counting rules~\cite{Manohar:1983md} for the potential
\bea
\mathcal{L}=\Lambda^2 f^2 \left(\frac{\Lambda}{4\pi f}\right)^{2L} \left(\frac{gf}{\Lambda}\right)^{2\eta} \left(\frac{\mu}{\Lambda}\right)^{\chi} \left(\frac{\pi}{f}\right)^{E_\pi},
\eea
where $\Lambda$ is the cutoff scale, $L=1$ is the loop order of the Higgs potential, $\mu$ is any masses or soft parameters, $\eta$ and $\chi$ are the number of insertions of the coupling $g$ and mass $\mu$, $E_\pi$ is the number of the Higgs fields.
Generally, it is showed the three-site model or models with more than three sites ($n\ge 3$) can typically guarantee the finite Higgs potential~\cite{Panico:2011pw}.

\section{Conclusions}
\label{sec:conclusion}

In this work, we present a warped five-dimensional new physics model, which can be regarded as one of the possible UV realizations of the minimal neutral naturalness model setup. Using the holographic method, we present the effective action on the UV brane which consists of only the holographic fields. Starting from the low energy effective Lagrangian for the top quark sector, we derive the finite Higgs potential with the correct vacuum misalignment angle, $\frac{v^2}{f^2}\ll 1$. Due to the soft breaking of the shift symmetry of the pseudo-Nambu-Goldstone boson, the Higgs potential is finite, i.e., without dependence on the cutoff scale. Comparing with the minimal neutral naturalness model, the cutoff dependence shown in the Higgs potential is replaced by the mass scale of the composite states. The small misalignment angle is strongly favored by the precision data of Higgs coupling measurements. Typically vacuum misalignment is realized through fermionic and bosonic cancellation. In this work, the vacuum misalignment is realized with only the fermionic contribution to the potential, which is the initial motivation that inspires us to construct concrete minimal neutral naturalness model. 
We deconstruct the holographic setup into four-dimensional composite Higgs model, and present the simplest two-site model to illustrate the main features in the composite sector.

In the minimal neutral naturalness model, due to the dark QCD group which is confined at around GeV scale, the smoking gun signature would be displaced vertices from the ``quirk" behavior at the colliders. In the UV completion, the new feature is the appearances of heavy composite states. We study the electrowek precision constraints on the scale of these composite states, which provides indirect probes of these states. On the other hand, the direct signatures of these composite states are worth to be explored at the LHC and future high energy hadron colliders. Finally the QCD color neutral composite states could have cosmological signatures, such as new dark matter candidates, etc.

\begin{acknowledgments} 
J.H.Y. is supported by the National Science Foundation of China under Grants No. 11875003 and the Chinese Academy of Sciences (CAS) Hundred-Talent Program.
L.X.X. and S.H.Z. are supported in part by the National Science Foundation of China under Grants No. 11635001, 11875072. 
\end{acknowledgments}


\appendix
\section{Fermion Profiles in Warped $AdS_5$}
With the general metric $ds^2\equiv a(z)^2\ (\eta_{\mu\nu}dx^\mu dx^\nu-dz^2)$, we have the equation of motion for the five-dimensional profile $f_{L,R}(p,z)$ of the bulk fermions $\Psi_{L,R}(p,z)$ as
\bea
\left(\partial_z+2\frac{\partial_za(z)}{a(z)}\pm a(z)M\right)f_{L,R}(p,z)=\pm pf_{R,L}(p,z)
\eea
where $M$ is the bulk mass and $p=\sqrt{p^2}$ and the profile is conveniently defined as
\bea
\Psi_{L,R}(p,z)=\frac{f_{L,R}(p,z)}{f_{L,R}(p,L_0)}\Psi^0_{L,R}(p).
\eea
We see $\Psi^0_{L,R}(p)$ is the corresponding field value on the UV brane.
For flat extra dimension, we can rewrite the equation of motion as
\bea
\left(\partial_z^2-M^2+p^2\right)f_{L,R}(p,z)=0
\eea
as $\partial_za(z)=0$. Solving the five-dimensional fermionic profile for the flat case would be straightforward~\cite{Serone:2009kf}.
For the case of warped $AdS_5$, we rewrite the equation of motion as
\bea
\left(\partial_z^2-\frac{4}{z}\partial_z+p^2+\frac{6}{z^2}\mp\frac{ML}{z^2}-\frac{(ML)^2}{z^2}\right)f_{L,R}(p,z)=0.
\eea
With IR-brane boundary conditions, we can solve the equation of motion and obtain all the five-dimensional profiles for bulk fermions~\cite{Contino:2004vy,Archer:2014qga,Croon:2015wba}.

In case there is mixing term on the IR brane for two bulk fermion fields $\Psi_1$ and $\Psi_2$ as
\bea
\mathcal{L}_{IR}=\frac{m}{g_5^2} \sqrt{-g_{\textrm{ind}}}\left(\bar{\Psi}_{1L}\Psi_{2R}+\bar{\Psi}_{2R}\Psi_{1L}\right) \ (z=L_1), 
\eea
we obtain the form factors for the holographic fields $\psi_L=\Psi_{1L}(z=L_0)$ and $\psi^\prime_R=\Psi_{2R}(z=L_0)$ defined as 
\bea
\bar{\psi}_L \frac{p\!\!\!/}{p} \Pi_L(m) \psi_L+\bar{\psi^\prime}_R \frac{p\!\!\!/}{p} \Pi_R(m) \psi^\prime_R+ \bar{\psi}_L \Pi_{LR}(m) \psi^\prime_R+\textrm{h.c.}\ .
\eea
We report the forms of $\Pi_L(m)$, $\Pi_R(m)$, and $\Pi_{LR}(m)$ as the following
\bea
\Pi_L(m)\equiv\frac{N_L(m)}{D_L(m)},\ \ \Pi_R(m)\equiv\frac{N_R(m)}{D_R(m)},\ \ \Pi_{LR}(m)\equiv\frac{N_{LR}(m)}{D_{LR}(m)}.
\eea
With the notations defined in Euclidean space for later convenience as
\begin{align}
G^{++}(\alpha_i,z)&= \left[K_{\alpha_i}(z Q)I_{\alpha_i}(L_1 Q)-K_{\alpha_i}(L_1 Q)I_{\alpha_i}(z Q)\right]\ ,\\
G^{+-}(\alpha_i,z) &= \left[K_{\alpha_i}(z Q)I_{\alpha_i-1}(L_1 Q)+K_{\alpha_i-1}(L_1 Q)I_{\alpha_i}(z Q)\right]\ ,\\
G^{-+}(\alpha_i,z) &= \left[K_{\alpha_i-1}(z Q)I_{\alpha_i}(L_1 Q)+K_{\alpha_i}(L_1 Q)I_{\alpha_i-1}(z Q)\right]\ ,\\
G^{--}(\alpha_i,z)  &= \left[K_{\alpha_i-1}(z Q)I_{\alpha_i-1}(L_1 Q)-K_{\alpha_i-1}(L_1 Q)I_{\alpha_i-1}(z Q)\right]\ ,\\
\tilde{G}^{+-}(\alpha_i,z) &= \left[K_{\alpha_i}(z Q)I_{\alpha_i-1}(z Q)+K_{\alpha_i-1}(z Q)I_{\alpha_i}(z Q)\right],
\end{align}
the rest of the form factors are 
\begin{align}
D_L(m)&=D_R(m)=\frac{D_{LR}(m)}{2}=m^2 G^{++}(\alpha_1, L_0) G^{--}(\alpha_2, L_0)+G^{-+}(\alpha_2, L_0) G^{+-}(\alpha_1, L_0)\ ,\\
N_L(m)&=i m^2 G^{-+}(\alpha_1, L_0) G^{--}(\alpha_2, L_0)+ i G^{-+}(\alpha_2, L_0) G^{--}(\alpha_1, L_0)\ ,\\
N_R(m)&=i m^2 G^{++}(\alpha_1, L_0) G^{+-}(\alpha_2, L_0)+ i G^{+-}(\alpha_1, L_0) G^{++}(\alpha_2, L_0)\ ,\\
N_{LR}(m)&= m \left\{ \tilde{G}^{+-}(\alpha_2,L_0) \tilde{G}^{+-}(\alpha_1,L_1)+ \tilde{G}^{+-}(\alpha_1,L_0) \tilde{G}^{+-}(\alpha_2,L_1)\right\}\ .
\end{align}
When the mixing parameter on the IR brane $m\to 0$, we see explicitly the form factors will reduce to
\bea
\Pi_{LR}(0)=0\ ,
\eea
\bea
\Pi_L(0)=i\ \frac{K_{\alpha_1-1}(L_0 Q)I_{\alpha_1-1}(L_1 Q)-K_{\alpha_1-1}(L_1 Q)I_{\alpha_1-1}(L_0 Q)}{K_{\alpha_1}(L_0 Q)I_{\alpha_1-1}(L_1 Q)+K_{\alpha_1-1}(L_1 Q)I_{\alpha_1}(L_0 Q)}\ ,
\eea
\bea
\Pi_R(0)=i\ \frac{K_{\alpha_2}(L_0 Q)I_{\alpha_2}(L_1 Q)-K_{\alpha_2}(L_1 Q)I_{\alpha_2}(L_0 Q)}{K_{\alpha_2}(L_1 Q)I_{\alpha_2-1}(L_0 Q)+K_{\alpha_2-1}(L_0 Q)I_{\alpha_2}(L_1 Q)}\ .
\eea
which is consistent with the results in Ref.~\cite{Contino:2004vy}.

\bibliographystyle{utphys}
\bibliography{minimal_long_final}

\end{document}